

\message
{JNL.TEX version 0.92 as of 6/9/87.  Report bugs and problems to Doug Eardley.}

\catcode`@=11
\expandafter\ifx\csname inp@t\endcsname\relax\let\inp@t=\input
\def\input#1 {\expandafter\ifx\csname #1IsLoaded\endcsname\relax
\inp@t#1%
\expandafter\def\csname #1IsLoaded\endcsname{(#1 was previously loaded)}
\else\message{\csname #1IsLoaded\endcsname}\fi}\fi
\catcode`@=12







\def\beginlinemode{\endmode
  \begingroup\parskip=0pt \obeylines\def\\{\par}\def\endmode{\par\endgroup}}
\def\beginparmode{\endmode
  \begingroup \def\endmode{\par\endgroup}}
\let\endmode=\par
{\obeylines\gdef\
{}}
\def\singlespace{\baselineskip=\normalbaselineskip}

\def\oneandahalfspace{\baselineskip=\normalbaselineskip
  \multiply\baselineskip by 3 \divide\baselineskip by 2}
\def\doublespace{\baselineskip=\normalbaselineskip \multiply\baselineskip by 2}






\def\title			
  {\null\vskip 3pt plus 0.2fill
   \beginlinemode \doublespace \raggedcenter \bf}

\def\author			
  {\vskip 3pt plus 0.2fill \beginlinemode
   \singlespace \raggedcenter\sc}

\def\affil			
  {\vskip 3pt plus 0.1fill \beginlinemode
   \oneandahalfspace \raggedcenter \sl}

\def\abstract			
  {\vskip 3pt plus 0.3fill \beginparmode
   \oneandahalfspace ABSTRACT: }

\def\endtitlepage		
  {\endpage			
   \body}

\def\body			
  {\beginparmode}		

\def\subhead#1{			
  \vskip 0.25truein		
  {\raggedcenter {#1} \par}
   \nobreak\vskip 0.25truein\nobreak}

\def\beginitems{
\par\medskip\bgroup\def\i##1 {\item{##1}}\def\ii##1 {\itemitem{##1}}
\leftskip=36pt\parskip=0pt}
\def\enditems{\par\egroup}

\def\beneathrel#1\under#2{\mathrel{\mathop{#2}\limits_{#1}}}

\def\refto#1{~[{#1}]}

\def\references			
  {
   \beginparmode
   \frenchspacing \parindent=0pt \leftskip=1truecm
   \parskip=8pt plus 3pt \everypar{\hangindent=\parindent}}

\gdef\refis#1{\item{#1.\ }}			

\gdef\journal#1, #2, #3, 1#4#5#6{		
    {\sl #1~}{\bf #2}, #3 (1#4#5#6)}		

\def\endreferences{\body}

\catcode`@=11
\newcount\r@fcount \r@fcount=0
\newcount\r@fcurr
\immediate\newwrite\reffile
\newif\ifr@ffile\r@ffilefalse
\def\w@rnwrite#1{\ifr@ffile\immediate\write\reffile{#1}\fi\message{#1}}

\def\writer@f#1>>{}
\def\referencefile{
  \r@ffiletrue\immediate\openout\reffile=\jobname.ref%
  \def\writer@f##1>>{\ifr@ffile\immediate\write\reffile%
    {\noexpand\refis{##1} = \csname r@fnum##1\endcsname = %
     \expandafter\expandafter\expandafter\strip@t\expandafter%
     \meaning\csname r@ftext\csname r@fnum##1\endcsname\endcsname}\fi}%
  \def\strip@t##1>>{}}

\def\citeall#1{\xdef#1##1{#1{\noexpand\cite{##1}}}}
\def\cite#1{\each@rg\citer@nge{#1}}	

\def\each@rg#1#2{{\let\thecsname=#1\expandafter\first@rg#2,\end,}}
\def\first@rg#1,{\thecsname{#1}\apply@rg}	
\def\apply@rg#1,{\ifx\end#1\let\next=\relax
\else,\thecsname{#1}\let\next=\apply@rg\fi\next}

\def\citer@nge#1{\citedor@nge#1-\end-}	
\def\citer@ngeat#1\end-{#1}
\def\citedor@nge#1-#2-{\ifx\end#2\r@featspace#1 
  \else\citel@@p{#1}{#2}\citer@ngeat\fi}	
\def\citel@@p#1#2{\ifnum#1>#2{\errmessage{Reference range #1-#2\space is bad.}%
    \errhelp{If you cite a series of references by the notation M-N, then M and
    N must be integers, and N must be greater than or equal to M.}}\else%
 {\count0=#1\count1=#2\advance\count1
by1\relax\expandafter\r@fcite\the\count0,%
  \loop\advance\count0 by1\relax
    \ifnum\count0<\count1,\expandafter\r@fcite\the\count0,%
  \repeat}\fi}

\def\r@featspace#1#2 {\r@fcite#1#2,}	
\def\r@fcite#1,{\ifuncit@d{#1}
    \newr@f{#1}%
    \expandafter\gdef\csname r@ftext\number\r@fcount\endcsname%
                     {\message{Reference #1 to be supplied.}%
                      \writer@f#1>>#1 to be supplied.\par}%
 \fi%
 \csname r@fnum#1\endcsname}
\def\ifuncit@d#1{\expandafter\ifx\csname r@fnum#1\endcsname\relax}%
\def\newr@f#1{\global\advance\r@fcount by1%
    \expandafter\xdef\csname r@fnum#1\endcsname{\number\r@fcount}}

\let\r@fis=\refis			
\def\refis#1#2#3\par{\ifuncit@d{#1}
   \newr@f{#1}%
   \w@rnwrite{Reference #1=\number\r@fcount\space is not cited up to now.}\fi%
  \expandafter\gdef\csname r@ftext\csname r@fnum#1\endcsname\endcsname%
  {\writer@f#1>>#2#3\par}}

\def\ignoreuncited{
   \def\refis##1##2##3\par{\ifuncit@d{##1}%
     \else\expandafter\gdef\csname r@ftext\csname
r@fnum##1\endcsname\endcsname%
     {\writer@f##1>>##2##3\par}\fi}}

\def\r@ferr{\endreferences\errmessage{I was expecting to see
\noexpand\endreferences before now;  I have inserted it here.}}
\let\r@ferences=\references
\def\references{\r@ferences\def\endmode{\r@ferr\par\endgroup}}

\let\endr@ferences=\endreferences
\def\endreferences{\r@fcurr=0
  {\loop\ifnum\r@fcurr<\r@fcount
    \advance\r@fcurr by 1\relax\expandafter\r@fis\expandafter{\number\r@fcurr}%
    \csname r@ftext\number\r@fcurr\endcsname%
  \repeat}\gdef\r@ferr{}\endr@ferences}


\let\r@fend=\endpaper\gdef\endpaper{\ifr@ffile
\immediate\write16{Cross References written on []\jobname.REF.}\fi\r@fend}

\catcode`@=12

\citeall\refto		


\magnification 1200
\vsize=7.8in
\hsize=5.3in
\voffset=0.1in
\hoffset=-0.05in
\newcount\eqnumber
\baselineskip 18pt plus 0pt minus 0pt


\font\rmmthree=cmbx10 scaled 1500
\font\rmmtwo=cmbx10 scaled 1200
\font\rmmoneB=cmbx10 scaled 1100

\font\rmmoneI=cmti10 scaled 1000

\font\ninerm=cmr10 scaled 900

\font\eightit=cmti10 scaled 800
\font\eightrm=cmr10 scaled 800
\font\eightbf=cmbx10 scaled 800


\def\title#1{\centerline{\noindent{\rmmthree #1}}\nobreak\medskip\eqnumber=1}

\def\sectbegin#1#2{\bigskip\bigbreak\leftline{\rmmtwo
#1~#2}\nobreak\medskip\nobreak}

\def\nosectbegin#1{\bigskip\bigbreak\leftline{\rmmtwo #1}\nobreak\medskip}

\def\subhead#1{\medskip\goodbreak\noindent{\bf #1}\nobreak\vskip 4pt\nobreak}


\def\lapp{\hbox{$ {     \lower.40ex\hbox{$<$}
                   \atop \raise.20ex\hbox{$\sim$}
                   }     $}  }
\def\gapp{\hbox{$ {     \lower.40ex\hbox{$>$}
                   \atop \raise.20ex\hbox{$\sim$}
                   }     $}  }

\def\marbul{\strut\vadjust{\kern-2pt$\bullet$}}

\def\rr{\rangle}
\def\ll{\langle}

\def\specialwarn{\vtop to
\strutdepth{\baselineskip\strutdepth\vss\llap{
\lower.1ex\hbox{$\bigtriangleup$}\kern-0.884em$\triangle$\kern-0.5667em{\eightrm
!}\hskip 13.5pt}\null}}
\def\strutdepth{\dp\strutbox}


\def\new{{\the\eqnumber}\global\advance\eqnumber by 1}
\def\delaynew{{\the\eqnumber}}
\def\nownew{\global\advance\eqnumber by 1}
\def\last{\advance\eqnumber by -1 {\the\eqnumber}
    \global\advance\eqnumber by 1}
\def\eqnam#1{
\xdef#1{\the\eqnumber}}


\def\figure#1#2#3#4#5{
\topinsert
\null
\medskip
\vskip #2\relax
\null\hskip #3\relax
\special{illustration #1}
\medskip
{\baselineskip 10pt\noindent\narrower\rm\hbox{\eightbf
Figure #4}:\quad\eightrm
#5 \smallskip}
\endinsert}

\def\vdoublefigure#1#2#3#4#5#6#7#8{
\topinsert
\medskip
\null
\vskip #3\relax
\null\hskip #4\relax
\special{illustration #1}
\smallskip
\vskip #6\relax
\line{\null\hskip #5\relax
\special{illustration #2}\hfill}
\medskip
{\baselineskip 10pt\noindent\narrower\rm\hbox{\eightbf
Figure #7}:\quad\eightrm
#8 \smallskip}
\endinsert}

\def\caption#1#2{
\baselineskip 10pt\noindent\narrower\rm\hbox{\eightbf
#1}:\quad\eightrm
#2 \smallskip}

\def\picture #1 by #2 (#3){
  \vbox to #2{
    \hrule width #1 height 0pt depth 0pt
    \vfill
    \special{picture #3} 
    }
  }

\def\scaledpicture #1 by #2 (#3 scaled #4){{
  \dimen0=#1 \dimen1=#2
  \divide\dimen0 by 1000 \multiply\dimen0 by #4
  \divide\dimen1 by 1000 \multiply\dimen1 by #4
  \picture \dimen0 by \dimen1 (#3 scaled #4)}
  }


\def\fa{f_{\rm a}}

\def\dalemb#1#2{{\vbox{\hrule height .#2pt
\hbox{\vrule width.#2pt height#1pt \kern#1pt\vrule width.#2pt}
\hrule height.#2pt}}}
\def\square{\mathord{\dalemb{5.9}{6}\hbox{\hskip1pt}}}
\def\tdot{\kern -8.5pt {}^{{}^{\hbox{...}}}}
\def\dotprime{\kern -8.0pt{}^{{}^{\hbox{.}~\prime}}}


\title{RADIATIVE BACKREACTION}
\title{ON GLOBAL STRINGS}
\vskip 20pt
\medskip
\centerline{\rmmoneB R.$\,$A. Battye {\rmmoneI~and~} E.$\,$P.$\,$S. Shellard}
\vskip 6pt
\baselineskip 20pt
\centerline{\rmmoneI Department of Applied Mathematics and Theoretical Physics}

\centerline{\rmmoneI University of Cambridge}

\centerline{{\rmmoneI Silver Street, Cambridge~CB3 9EW, U.K.}
\footnote{*}{{\noindent Email: rab17$\,$@$\,$damtp.cam.ac.uk and
epss$\,$@$\,$damtp.cam.ac.uk}
\smallskip\indent Paper
submitted to {\eightit Physical Review D}.\smallskip}}

\vskip 12pt
\bigskip \centerline{\rmmoneB Abstract} \medskip {\narrower{\baselineskip 16pt
\ninerm \noindent We consider radiative backreaction for global
strings using the Kalb-Ramond formalism. In analogy to the point electron
in classical electrodynamics, we show how local radiative corrections to the
equations of motion allow one to remove the divergence in the self field and
calculate a first order approximation to the radiation backreaction force. The
effects of this backreaction force are studied numerically by
resubstituting the equations of motion to suppress exponentially growing
solutions. By direct comparison with numerical field theory
simulations and analytic radiation calculations we establish that the `local
backreaction approximation' provides a satisfactory quantitative
description of radiative damping for a wide variety of string configurations.
Finally, we discuss the relevance of this work to the evolution of a network of
global strings and their possible cosmological consequences. These methods can
also be applied to describe the effects of gravitational radiation backreaction
on local strings, electromagnetic radiation backreaction on superconducting
strings and other forms of string radiative backreaction. \smallskip}}   \vskip
20pt \smallskip \baselineskip 20pt\tenrm

\sectbegin{1.}{Introduction}

\noindent A variety of unified field theories predict the formation
of a network of topological defects at one or more phase
transitions in the early universe\refto{kibble}. Strings associated with the
breaking of local symmetries have generated the most interest in the
literature because, amongst other reasons, GUT-scale strings could have been
the
initial seeds for the formation of large-scale structure\refto{structure}.
However, local strings are tightly constrained by their contribution to the
gravitational radiation background\refto{AC}.  There are other types of strings
which circumvent this constraint and which may have similar cosmological
implications, in particular those formed when a global symmetry is broken.
Instead of radiating gravitationally, the dominant radiation mechanism for
these
strings is the emission of massless Nambu-Goldstone bosons\refto{DavVV}. In a
recent publication\refto{BSa}, we studied the nature of this radiation in
detail, using analytic and numerical techniques. We demonstrated that a low
energy effective action known as the Kalb-Ramond action, provided an accurate
description of the dynamics of global strings even at the moderately high
velocities one expects in a realistic string network. Within this formalism the
topological coupling of the massless field to the string is linearized. One
finds that the coupling between the field and the string worldsheet is similar
to that of the point electron in electromagnetism.  However, there are still
difficulties associated with this approach, notably because equations of motion
are inconsistent due to a divergent self-field.

This type of problem has been well understood for some time in the context of
a point electron in classical electrodynamics\refto{dirac}. In
the case of the electron, the self and radiation fields can be distinguished
easily since, at large distances $R$, the self field falls off as $1/R^2$,
whereas the radiation field falls off as $1/R$. Careful analysis of the
equations of motion leads to the renormalisation of the electron mass by the
Coulomb self-field, using the classical electron radius to cut off short
distance divergences and a first order approximation to the radiation
backreaction force known as the Abraham-Lorentz force,  \eqnam{\abrabamlorentz}
$$F_{\mu}^{\rm rad}= {2\over 3}~{e^2\over 4\pi}\bigg{(}X^{\tdot}_{\mu}+\ddot
X^2 \dot X_{\mu}\bigg{)}\,,\eqno(\new)$$ where $X_{\mu}(\tau)$ is the position
on the electron's worldline at time $\tau$. The dependence of this force on
$X^{\tdot}_{\mu}$ has lead to problems in numerical applications since there
exist exponentially increasing solutions to the equations of motion. These
unphysical `runaway' solutions can only be suppressed by rewriting the
equations of motion as an integro-differential equation.

In a recent letter\refto{BSc}, we proposed a formalism for removing the self
force and calculating a first approximation for the radiation backreaction
force
of strings using the analogy of classical electrodynamics. However, the
analogy is not exact because the strings are line-like objects of possibly
infinite extent. We circumvent this problem by assuming that the dominant
contribution to the backreaction force comes from string segments in the
vicinity of the point in question, henceforth known  as the `local
backreaction approximation'. This approximation will not be valid in
every situation, but in a wide variety of circumstances it should work
well.  In this paper we will consider this
approximation in greater detail. We will elaborate on the derivation of the
radiation backreaction force and give a discussion of the physical aspects of
the approximation. Further numerical evidence will be presented in support of
the validity of this approximation in physically important cases by direct
comparison between modified Nambu dynamics, evolved numerically using the
backreaction force, and numerical field theory simulations.  The one free
parameter in our analysis, effectively the damping coefficient, can be
normalized by comparing with numerical field theory simulations and known
analytic results.

One of the main motivations for this work is to implement appropriate radiative
corrections in a full network simulation. We anticipate that the scaling
assumption for gauge strings, numerically verified in refs.\refto{BB,ASa},
will also be seen to be valid for global strings. However, it is anticipated
that the parameters quantifying the small scale features will be somewhat
different\refto{BSb}. Accurate numerical simulations will allow estimates of
the
cosmological axion density to be refined. A similar formalism is applicable
to gravitational radiation backreaction on local strings\refto{Bata} and
electromagnetic radiation backreaction on superconducting strings.

Throughout this paper we employ a (+$\,---$) signature for the spacetime metric
$g_{\mu\nu}$ and (+$\,-$) for the induced metric on the string worldsheet
$\gamma_{ab}$, the coordinates for which are given by $X^{\mu} =
X^{\mu}(\sigma ,\tau)$, with the null coordinates,
$u=\sigma-\tau, ~v=\sigma+\tau$.

\sectbegin{2.}{Analytic formalism}

\subhead{2.1 The Kalb-Ramond action}

\noindent Th essential features of global strings in flat space are exhibited
in the simple U(1) Goldstone model, with action given by
\eqnam{\goldaction}
$$ S=\int d^4x\bigg{[}\partial_{\mu}\bar\Phi\partial^{\mu}\Phi-{1\over
4}\lambda(\bar\Phi\Phi-\fa^2)^2\bigg{]}\,,\eqno(\new)$$
where $\Phi=\phi e^{i\vartheta}$ is a
complex scalar field which can be split into a massive (real) component $\phi$
and a massless (real periodic) Goldstone boson $\vartheta$. The analytic
treatment of global string dynamics is hampered by the topological coupling of
the self field of the string to the Goldstone boson radiation field. However,
we can exploit the well-known duality between a massless scalar field and a
two-index antisymmetric tensor $B_{\mu\nu}$ to replace the Goldstone
boson $\vartheta$ in (\goldaction) via the relation
\eqnam{\duality}
$$\phi^2\partial_{\mu}\vartheta = {1\over
2}\fa\epsilon_{\mu\nu\lambda\rho}\partial^{\nu}B^{\lambda\rho}\,.\eqno(\new)$$
Performing this transformation carefully and integrating over the massive
degrees of freedom about the two-dimensional string worldsheet $X^\mu(\sigma,
\tau)$\refto{DSb},  yields the flat-space Kalb--Ramond
action\refto{KR,VV},\eqnam{\kraction}
$$ S=-\mu_0\int d\sigma d\tau\sqrt{-\gamma}+{1\over 6}\int d^4x H^2 -
2\pi\fa\int B_{\mu\nu}d\sigma^{\mu\nu}\,,\eqno(\new)$$
where $H_{\mu\alpha\beta}=\partial_{\mu}B_{\alpha\beta}+
\partial_{\beta}B_{\mu\alpha}+\partial_{\alpha}B_{\beta\mu}$ is the field
strength of the antisymmetric tensor field $B_{\mu\nu}$, the metric induced
on the world sheet is
$$\gamma_{ab}=g_{\mu\nu}\partial_aX^{\mu}\partial_bX^{\nu}\,,\quad\gamma={\rm
det}(\gamma_{ab})\,,\eqno(\new)$$ and the area element on the worldsheet is
$$d\sigma^{\mu\nu}=\epsilon^{ab}\partial_aX^{\mu}\partial_bX^{\nu}d\sigma
d\tau\,.$$ The first term is the familar Nambu action for local strings, the
second is the antisymmetric field strength for both external fields and the
self-field of the string and the last term is a contact interaction between
the antisymmetric tensor field and the string worldsheet. The coupling
between the string and the antisymmetric tensor is analogous to the
electromagnetic coupling of the point electron to the electromagnetic
field. This analogy underpins our subsequent development of global string
dynamics based on (\kraction).

Varying the action (\kraction) with respect to the worldsheet coordinates and
the antisymmetric tensor yields the string equations of motion and the tensor
field equations,
\eqnam{\aceom}
$$\eqalign{&\mu_0\partial_a(\sqrt{-\gamma}\gamma^{ab}\partial_b
X^{\mu})={\cal F}^{\mu}=2\pi\fa
H^{\mu\alpha\beta}V_{\alpha\beta}\,,\cr&\partial_{\mu}H^{\mu\alpha\beta}=-4\pi
J^{\alpha\beta}=-2\pi\fa\int d\sigma
d\tau\delta^4\big{(}x-X(\sigma,\tau)\big{)}V^{\alpha\beta}\,, }\eqno(\new)$$
where $V_{\alpha\beta}=\epsilon^{ab}\partial_aX_{\alpha}\partial_bX_{\beta}$
is the antisymmetric vertex operator. In the conformal string gauge and the
Lorentz antisymmetric tensor gauge,
\eqnam{\gaugecon}
$$\dot X^2+X^{\prime 2}=0\,,\quad \dot X\cdot X^{\prime}=0\,,\quad
\partial_{\mu}B^{\mu\nu}=0\,,\eqno(\new)$$
the equations of motion (\aceom) become
\eqnam{\modeom}
$$\eqalign{&\mu_0\big{(}\ddot
X^{\mu}-X^{\prime\prime\mu}\big{)}={\cal F}^{\mu}=2\pi\fa
H^{\mu\alpha\beta}V_{\alpha\beta}\,,\cr&\square B_{\alpha\beta}=-4\pi
J_{\alpha\beta}=-2\pi\fa\int d\sigma
d\tau\delta^4\big{(}x-X(\sigma,\tau)\big{)}V_{\alpha\beta}\,,}\eqno(\new)$$
where $\square=g^{\mu\nu}\partial_{\mu}\partial_{\nu}$ and
$V_{\alpha\beta}=\dot
X_{\alpha}X^{\prime}_{\beta}-X^{\prime}_{\alpha}\dot X_{\beta}$.  These
equations are problematic because the self-field diverges as any point of the
string is approached, that is $x\rightarrow X(\sigma,\tau)$.

\subhead{2.2 Simple string configurations}

\noindent If one ignores the effects of the force density ${\cal F}^{\mu}$,
then
the equations of motion reduce to the well-known Nambu equations of motion, a
massless wave equation. The equations have solution
\eqnam{\nambusoln}
$$X^{0}=t=\tau\,,\quad {\bf X}={1\over 2}\big{[}{\bf a}(u)+{\bf
b}(v)\big{]}\,,\eqno(\new)$$
where the functions ${\bf a}(u)$ and ${\bf b}(v)$ are the left- and
right-moving parts of the solution (recall $u=\sigma-t$ and $v=\sigma+t$).
Using
the conformal gauge conditions (\gaugecon), one can deduce that
\eqnam{\abdef}
$${\bf a}^{\prime 2}=1\,,\quad {\bf b}^{\prime 2}=1\,.\eqno(\new)$$

\figure{loop_long.eps}{2.0in}{-0.45in}{1}{Schematic of the solutions of the
Nambu equations of motion: (a) loop solutions parametrized by the
invariant length $L$ and (b) long string solutions parameterized by the
wavelength $L$ and the relative amplitude ${\cal E}$.}

The equations (\nambusoln) and (\abdef) have
closed loop and periodic long (or infinite) string solutions. The
loop solutions are parametrized by the length of the loop $L$, which is
closely related to the characteristic frequency  $\Omega=2\pi/L$, whereas the
long periodic solutions are parametrized by the wavelength $L$ and the ratio
of amplitude to wavelength or the relative amplitude ${\cal E}=2\pi A/L$,
where $A$ is the amplitude. Fig.~1 shows a schematic of the two types of
solution we shall consider. In general situations, such solutions will
correspond to a superposition of a large number of harmonics.

A simple two parameter family of loops, known as Kibble-Turok loops,
involve just the first and third harmonics\refto{kibbturok}:
\eqnam{\kibbleturoksolndef}
$$
\eqalign{{\bf X} = &{1\over
2\Omega}\bigg{(}(1-\alpha)\sin\Omega u  +\textstyle{1\over 3}
\alpha\sin3\Omega u + \sin\Omega v,\cr
&\quad-(1-\alpha)\cos\Omega u-
\textstyle{1\over 3}\alpha\cos 3\Omega u -\cos\psi\cos\Omega v,\cr
&\quad
-2\big{[}\alpha(1-\alpha)\big{]}^{1/2}\cos\Omega u-\sin\psi\cos\Omega
v\bigg{)}\,,}
\eqno(\new)
$$
where $\Omega=2\pi/L$ and
$0\le\alpha\le1,-\pi\le\psi\le\pi$. If $\alpha=0$ and $\psi=0$ then the
solution is a circular loop, which oscillates between a circle of radius
$L/2\pi$ and a point. For a significant range of the parameters $\alpha$
and $\psi$ these solutions can be shown to be non self-intersecting and
so the dominant decay mechanism is likely to be through radiation rather
than loop fragmentation. The time evolution of a particular solution with
$\psi=\pi/3$ and $\alpha=0.5$ is shown in fig.~2a.
Kibble--Turok loops also generate cusps, that is, points on the string which
reach the velocity of light. For example, cusps will appear on the $\alpha=0$
solution at  $\sigma=L/4,\,3L/4$ when $t=(n+1/2)L/2$ ($n$
integer). Whether or not cusps are generic on realistic loops has been the
subject of various heuristic discussion, which have also considered the unknown
effect of backreaction on cusp evolution.

Strings reconnect or `exchange partners' when they intersect.  This process
introduces kinks---contact discontinuities in the velocity $\dot{\bf  X}$ and
tangent vector ${\bf X}'$---which propagate along the string at the speed of
light.  Because realistic loops are produced by long string reconnections or
self-intersections we can be sure they will possess at least two kinks,
probably more.  An idealized loop with four kinks between four straight
string segments is given by the following\refto{GV} \eqnam{\kinkyloop}
$${\bf X}={1\over 2}\big{[}{\bf a}(\sigma-\tau)+{\bf
b}(\sigma+\tau)\big{]}\,,$$
where
\eqnam {\akink}
$${\bf a}(\sigma-\tau) = \left\{\eqalign{
&\bigg{(}{L\over 2\pi}(\sigma-\tau)-{L\over 4}\bigg{)}{\bf A}\cr &
\bigg{(}{3L\over 4}-{L\over 2\pi}(\sigma-\tau)\bigg{)}{\bf A}}
\right.\hbox{      } \eqalign{&0<\sigma-\tau<\pi \cr
&\pi<\sigma-\tau<2\pi\,,} \eqno(\new)$$ \eqnam{\bkink}
$${\bf b}(\sigma+\tau) = \left\{\eqalign{
&\bigg{(}{L\over 2\pi}(\sigma+\tau)-{L\over 4}\bigg{)}{\bf B} \cr &\bigg{(}
{3L\over 4}-{L\over 2\pi}(\sigma+\tau)\bigg{)}{\bf B}} \right.\hbox{      }
\eqalign{&0<\sigma+\tau<\pi \cr &\pi<\sigma+\tau<2\pi\,,}
\eqno(\new)$$
with  arbitrary unit vectors ${\bf A}$ and ${\bf B}$. The two pairs of kinks
propagate in opposite directions around the loop. In the special case ${\bf
A}\cdot{\bf B}$, the loop is planar and oscillates between a square and a
doubled line.

A simple, symmetric long string  solution can be constructed from equal
and oppositely propagating helicoidal waves in the fundamental
mode\refto{Sak}, \eqnam{\helixsolndef}
$$ \eqalign{{\bf X} &=
\bigg{(}{{\cal E} \over 2\Omega}\big{[}\cos\Omega u + \cos\Omega v
\big{]},{{\cal E} \over 2\Omega}\big{[}\sin\Omega u + \sin\Omega v \big{]}, {1
\over 2}\sqrt{1-{\cal E}^2}\,(u+v)\bigg{)}\cr &=
\bigg{(}{{\cal E}\over\Omega}\sin\Omega\sigma\cos\Omega
t,{{\cal E}\over\Omega} \cos\Omega\sigma\cos\Omega
t,\sqrt{1-{\cal E}^2}\sigma\bigg{)}\,,}\eqno(\new)$$ where $0<{\cal E} <1$
and ${\cal E} \to 1$ in the relativistic limit. This  corresponds to a
helicoidal solution which oscillates between a static helix and a  straight
line, as shown in fig.~2(b) for ${\cal E}=0.6$. Because of its perfect
symmetry, calculations for the radiation power from this solution are
analytically  tractable\refto{Sak,BSa}, though such configurations are unlikely
to be found in a realistic string network.

One can generalise the helicoidal solution (\helixsolndef) to have
unequal left and right moving amplitudes, \eqnam{\leftrightsolndef}
$$
{\bf X} =
\bigg{(}{{\cal E}_R \over 2\Omega}\cos\Omega u +{{\cal E}_L \over
2\Omega}\cos\Omega u  ,{{\cal E}_R \over 2\Omega}\sin\Omega u + {{\cal E}_L
\over 2\Omega}\sin\Omega u , {1 \over
2}\sqrt{1-{\cal E}_R^2}\,u+{1 \over
2}\sqrt{1-{\cal E}_L^2}v\bigg{)}\,,\eqno(\new)
$$
where
$0<{\cal E}_R,{\cal E}_L <1$. This type of solution is thought to be a
reasonably accurate description of long strings in a realistic network, since
within a sufficiently small volume the number of left and right movers are
unlikely to be strongly correlated. One special case of
this solution is that with no left moving perturbation, \eqnam{\rightsolndef}
$$
{\bf X} = \bigg{(}{{\cal E}_R \over 2\Omega}\cos\Omega u ,{{\cal E}_R \over
2\Omega}\sin\Omega u , {1 \over 2}\sqrt{1-{\cal E}_R^2}\,u+{1\over
2}v\bigg{)}\,.\eqno(\new)
$$
It has been suggested that pure left- or right-moving configurations do not
radiate and so they will propagate indefinitely\refto{Vac}.  However, we shall
argue that such solutions are not physically relevant because they require
initial data with the string fields artificially correlated
out to infinity.

A solution, similar to the helix (\helixsolndef), but with sinusoidal
perturbations in only one plane is \eqnam{\cosinesolndef}
$$
{\bf X} = \bigg{(}{{\cal E} \over
2\Omega}\big{[}\cos\Omega u + \cos\Omega v  \big{]},0,{1 \over
2\Omega}\big{[}E({\cal E},\Omega u) + E({\cal E},\Omega v)
\big{]}\bigg{)}\,,\eqno(\new)
$$
where $E(k,\phi)$ is the incomplete
elliptic integral of the second kind, defined by
\eqnam{\ellipicdef}
$$
E(k,\phi)=\int_0^{\phi}d\theta\sqrt{1-k^2\sin
^2\theta}\,,\eqno(\new)
$$
where  $0<{\cal E} <1$ and ${\cal E} \to 1$ in
the relativistic limit.

Long string solutions of (\abdef) can also possess kinks as in (\akink). A
simple solution for a periodic distribution of kinks on a perturbed
straight string, consists of the left- and right-moving perturbations, $X_U$
and
$X_V$  respectively,
\eqnam{\kinksolndef}
$$
{\bf X} = \bigg{(}X_U+X_V,0,{1 \over 2}\sqrt{1-{4{\cal E}^2 \over \pi^2}}(u+v)
\bigg{)}\,,\eqno(\new)
$$
\noindent where
\eqnam {\kinkone}
$$
X_U = \left\{\eqalign{
&{2{\cal E}  \over \pi}u \cr &{2{\cal E}  \over \pi}\bigg{(}{1 \over 2}L-u
\bigg{)} \cr
&{2{\cal E}  \over \pi}\bigg{(}-L+u\bigg{)}} \right.\hbox{  }
\eqalign{&0<u<{1 \over 4}L \cr &{1 \over 4}L<u<{3 \over 4}L \cr&{3 \over
4}L<u<L\,,}
\eqno(\new)
$$
\eqnam{\kintwo}
$$
X_V = \left\{\eqalign{
&{2{\cal E}  \over \pi}v \cr &{2{\cal E}  \over \pi}\bigg{(}{1 \over 2}L-v
\bigg{)} \cr
&{2{\cal E}  \over \pi}\bigg{(}-L+v\bigg{)}} \right.\hbox{  }
\eqalign{&0<v<{1 \over 4}L \cr &{1 \over 4}L<v<{3 \over 4}L \cr&{3 \over
4}L<v<L\,,}
\eqno(\new)
$$
\noindent where $0<{\cal E} < {\pi \over 2}$ and, in this case,
${\cal E}\to{\pi\over 2}$ is the relativistic limit. In fig.~2d, notice
how the kinks split into two which propagate at the speed of light in
opposite directions along the string.

\figure{large1.eps}{4.0in}{0.5in}{2}{The evolution of various string
trajectories: (a) Kibble-Turok loop (\kibbleturoksolndef)
($\psi=\pi/3$, $\alpha=0.5$), (b) four kink loop solution (\akink) (c)
helicoidal
long string (\helixsolndef) with $({\cal E}=0.6)$, and (d) kink solution
(\kinksolndef) with $45^{\circ}$ openning angle.}

\subhead{2.3 Green functions}

\noindent One of the most basic techniques of mathematical physics is the
inversion of differential equations such as the field equation in
(\modeom) using Green functions. The basic Green functions satisfy
\eqnam{\defgreen}
$$\square D(x)=\delta^4(x)\,,\eqno(\new)$$ which implies that the
solution to   \eqnam{\defeqn}
$$\square F(x)=S(x)\,,\eqno(\new)$$ is given by integrating the product of
the Green function and the forcing term $S(x)$ over spacetime
\eqnam{\answer}
$$F(x)=\int d^4x^{\prime} D(x-x^{\prime})S(x^{\prime})\,.\eqno(\new)$$

In order to deduce a specific form for the Green function one must specify some
boundary conditions which define the region over which initial data is known.
The two most common Green functions used are the retarded and advanced time
Green functions, which use initial data on the backward or forward light
cones respectively, \eqnam{\retadv} $$D_{\rm ret}(x)={1\over
2\pi}\theta(x^0)\delta(x^2)\,,\quad D_{\rm adv}(x)={1\over
2\pi}\theta(-x^0)\delta(x^2)\,,\eqno(\new)$$ where $x^{\mu}=(x^0,{\bf x})$
and $\theta(x^0)$ is the Heaviside function, that is $\theta(x^0)=1$ for
$x^0>0$ and $\theta(x^0)=0$  otherwise.

In problems where radiation is involved one wishes to separate radiative
effects from those of the self-field. The radiation field is free and the
radiation Green function  must satisfy a homogeneous version of
(\defgreen). One can construct such a Green function by subtracting the
advanced Green function from the retarded. Similarly, one can construct the
Green function for the self-field by summing the retarded and advanced
Green's functions. Using appropriate normalisations, one can deduce that
\eqnam{\selfradgreen} $$\eqalign{D_{\rm rad}&={1\over 2}\bigg{(}D_{\rm
ret}-D_{\rm adv}\bigg{)} = {1\over 4\pi}\epsilon(x^0)\delta(x^2)\,,\cr
D_{\rm self}&={1\over 2}\bigg{(}D_{\rm ret}+D_{\rm adv}\bigg{)}= {1\over
4\pi}\delta(x^2)\,,}\eqno(\new)$$ where
$\epsilon(x^0)=\theta(x^0)-\theta(-x^0)$. One can calculate the  self- and
radiation-field for a problem such as (\defeqn) using similar expression to
(\answer) with the basic Green function replaced by the appropriate
expression from (\selfradgreen).

\subhead{2.4 Lienard-Wiechart Potentials}

\noindent Using the Green function techniques described in the previous
section, one can deduce that  \eqnam{\bsoln}
$$\eqalign{B_{\alpha\beta}(x)& = -4\pi\int d^4x^{\prime}D_{\rm
ret}(x-x^{\prime})J_{\alpha\beta}(x^{\prime}) \cr &= -2\pi\fa\int d\bar\sigma
d\bar\tau \, D_{\rm
ret}(x-X(\bar\sigma,\bar\tau))V_{\alpha\beta}(\bar\sigma,\bar\tau)\,.}\eqno(\new)$$
The integration is over all time and over all string segments. In the case of
a closed loop this is a finite range, but for long (infinite) strings
the range is infinite. If one
defines $\Delta_{\mu}=x_{\mu}-X_{\mu}(\bar\sigma,\bar\tau)$ while treating
$\bar \sigma$ as $\bar\sigma=\bar\sigma(\bar\tau)$, then
\eqnam{\deltadef}
$$
\eqalign{&d(\Delta^2)=-2\Delta\cdot\dot X d\bar\tau\,,\cr
&\partial_{\rho}=2\Delta_{\rho}{\partial\over\partial(\Delta^2)}=-{\Delta_{\rho}\over
\Delta\cdot\dot X}{\partial\over\partial\bar\tau}\,.}
$$
Substituting into (\bsoln) and evaluating the delta function one can deduce
the Lienard-Wiechart potential\refto{dirac,DQ,HCH}
\eqnam{\lwpot} $$
B_{\alpha\beta}(x)= -{\fa\over 2}\int d\bar\sigma
\bigg{(}{V_{\alpha\beta}\over |\Delta.\dot
X|}\bigg{)}\bigg{|}_{\bar\tau=\tau_R}\,,\eqno(\new)$$  where
$\Delta^2|_{\bar\tau=\tau_R}=0$ and $\tau_R<t$. The modulus sign in
(\lwpot) preserves the orientation of the region of integration when
evaluating the delta-function.
In order to calculate the radiation backreaction force one requires the
derivative of (\lwpot). This can be calculated by performing an integration
by parts, \eqnam{\lwpotderiv} $$ \partial_{\mu}B_{\alpha\beta}(x)=-
{\fa\over 2}\int d\bar\sigma  {1\over \Delta.\dot X}{\partial\over
\partial\bar\tau}\bigg{(}{\Delta_{\mu}V_{\alpha\beta}\over |\Delta.\dot
X|}\bigg{)}\bigg{|}_{\bar\tau=\tau_R}\,.\eqno(\new) $$

One can separate the
radiation field from the self field by using the Green functions for the
self- and radiation-fields (\selfradgreen). Therefore, one can calculate the
Lienard-Wiechart potentials and their derivatives for both the
self and radiation fields. \eqnam{\lwpotselfrad}  $$
\eqalign{B^{\rm self}_{\alpha\beta}(x)&=-{\fa\over 4}\int
d\bar\sigma\bigg{[}{V_{\alpha\beta}\over |\Delta\cdot\dot
X|}\bigg{|}_{\bar\tau=\tau_R} +{V_{\alpha\beta}\over |\Delta\cdot\dot
X|}\bigg{|}_{\bar\tau=\tau_R^{\prime}}\bigg{]}\,,\cr B^{\rm
rad}_{\alpha\beta}(x)&=-{\fa\over 4}\int
d\bar\sigma\bigg{[}{V_{\alpha\beta}\over |\Delta\cdot\dot
X|}\bigg{|}_{\bar\tau=\tau_R} -{V_{\alpha\beta}\over |\Delta\cdot\dot
X|}\bigg{|}_{\bar\tau=\tau_R^{\prime}}\bigg{]}\,,\cr\partial_{\mu}B_{\alpha\beta}^{\rm
self}(x)&=-{\fa\over 4}\int d\bar\sigma  \bigg{[}{1\over \Delta\cdot\dot X}
{\partial\over \partial\bar\tau}\bigg{(}{\Delta_{\mu}V_{\alpha\beta}\over
|\Delta\cdot\dot X|}\bigg{)}\bigg{|}_{\bar\tau=\tau_R}+{1\over \Delta\cdot\dot
X}
{\partial\over \partial\bar\tau}\bigg{(}{\Delta_{\mu}V_{\alpha\beta}\over
|\Delta\cdot\dot X|}\bigg{)}\bigg{|}_{\bar\tau=\tau^{\prime}_R}\bigg{]}\,,\cr
\partial_{\mu}B_{\alpha\beta}^{\rm rad}(x)&=-{\fa\over 4}\int d\bar\sigma
\bigg{[}{1\over \Delta\cdot\dot X} {\partial\over
\partial\bar\tau}\bigg{(}{\Delta_{\mu}V_{\alpha\beta}\over |\Delta\cdot\dot
X|}\bigg{)}\bigg{|}_{\bar\tau=\tau_R}-{1\over \Delta\cdot\dot X} {\partial\over
\partial\bar\tau}\bigg{(}{\Delta_{\mu}V_{\alpha\beta}\over |\Delta\cdot\dot
X|}\bigg{)}\bigg{|}_{\bar\tau=\tau^{\prime}_R}\bigg{]}\,. }\eqno(\new) $$
where $\Delta^2|_{\bar\tau=\tau_R,\tau^{\prime}_R}=0$ and
$\tau_R<t,\tau^{\prime}_R>t$.  Effectively, then, we have performed the
separation of the self- and radiation-fields. An attempt was made to perform
this split in ref.\refto{VacFort}, using techniques similar to those used in
classical electrodynamics\refto{dirac}. This method  performed the split on the
basis of asymptotic fall off. As already discussed, this procedure works in
the case of the electron since the self-field falls off like $1/R^2$, whereas
the radiation field falls off like $1/R$ for large $R$.  However, this
procedure may work for string loops. However, it is
doomed to failure for long strings, since both the self- and radiation-fields
fall off as $1/R$.

\subhead{2.5 The `local backreaction approximation'}

\noindent It has already been noted that the renormalisation procedure for
strings is more complicated than that for the point electron. The main problem
becomes obvious when one compares the Lienard-Wiechart potentials for strings
to those for the electron\refto{dirac}. Since the string is an extended object,
the Lienard-Wiechart potential is an integral along the string. In the case of
a
loop of length $L$, this integral will be in the range
$0<|\sigma-\bar\sigma|<L$
for a point $X(\sigma,\tau)$ on the string and can be easily approximated.
However, in the case of a long (infinite) string  the range is
$-\infty<|\sigma-\bar\sigma|<\infty$ and the integral cannot be evaluated
without the solution being  periodic.

In more general situations this is not possible and one must make what we shall
call the `local backreaction approximation'. Since the effects of backreaction
from string segments at large distances from the point in question must be
suppressed, it seems sensible to truncate the integrals of (\lwpot) and
(\lwpotselfrad) at some renormalisation scale $\Delta$, which is at present
arbitrary. That is the integrals are over the range
$-\Delta/2<|\sigma-\bar\sigma|<\Delta/2$. Using the case of the loop of length
$L$ as an example seems to suggest that $\Delta\sim L$. In fact our expectation
is that, for more general string trajectories, an appropriate choice for
$\Delta$
would be near the average curvature radius of the string.

Using the local backreaction approximation one can perform the renormalisation
of the self-field and the derivation of a first order approximation to the
radiation-field. If one allows $x\to X(\sigma,\tau)$, then (\lwpotselfrad)
can be expanded in terms of  $s=\sigma-\bar\sigma$ and $t=\tau-\bar\tau$. This
procedure requires that the natural scale for the otherwise arbitrary
renormalisation cut-off $\Delta$ be less than the average curvature radius of
the string. In this case one finds that the condition $\Delta^2|_{\tau_R}=0$,
implies that $t=|s|+O(s^4)$ and $\Delta^2|_{\tau_R^{\prime}}=0$, implies that
$t=-|s|+O(s^4)$. Ignoring terms of order four in $s$ and $t$, allows one to
deduce that  \eqnam{\renfieldsconf} $$\eqalign{H_{\mu\alpha\beta}^{\rm
self}=&{\fa\over 2\dot X^4}\bigg{[}\ddot X_{[\mu}V_{\alpha\beta]} -
X^{\prime\prime}_{[\mu}V_{\alpha\beta]}\bigg{]}\log(\Delta/\delta)+{\cal O}
(\Delta^2)\,,\cr H_{\mu\alpha\beta}^{\rm rad}=&{\fa\over 2\dot
X^4}\bigg{[}-{4\over 3}X^{\tdot}_{[\mu}V_{\alpha\beta]}-{1\over 2}\dot
X_{[\mu}\dot V_{\alpha\beta]}+\bigg{(} {2\dot X.\ddot X\over \dot
X^2}\bigg{)}\ddot X_{[\mu}V_{\alpha\beta]}\bigg{]}\Delta+{\cal O}
(\Delta^2)\,,\cr}\eqno(\new) $$ where
$A_{[\mu\alpha\beta]}=A_{\mu\alpha\beta}+A_{\beta\mu\alpha}+A_{\alpha\beta\mu}$.
Note that the self-field has no order $\Delta$ term. Ignoring terms of order
$\Delta^2$, we can then obtain expressions for the self-force and the
first order approximation to the radiation backreaction force density,
\eqnam{\backforcetwo}
$$
\eqalign{{\cal F}^{\rm
self}_{\mu}=&-2\pi\fa^2\log(\Delta/\delta)\bigg{[}\ddot X_{\mu}
-X^{\prime\prime}_{\mu}\bigg{]} \cr
{\cal F}^{\rm rad}_{\mu}=&
\pi\fa^2\Delta\bigg{\{}{4\over 3}X^{\tdot}_{\mu} -2 \bigg{(}{\dot X.\ddot
X\over
\dot X^2}\bigg{)}\ddot X_{\mu} +2 \bigg{(}{X^{\prime}.\ddot X\over \dot
X^2}\bigg{)} X_{\mu}^{\dotprime} +\bigg{[}-{4\over 3}\bigg{(}{\dot
X.X^{\tdot}\over \dot X^2}\bigg{)} +2\bigg{(}{\dot X.\ddot X\over \dot
X^2}\bigg{)}^2  \cr
&\qquad +2\bigg{(}{X^{\prime}.\ddot X\over\dot
X^2}\bigg{)}^2\bigg{]}\dot X_{\mu} +\bigg{[}{4\over
3}\bigg{(}{X^{\prime}.X^{\tdot}\over \dot X^2} \bigg{)} - 4\bigg{(}{(\dot
X.\ddot X)(X^{\prime}\ddot X)\over \dot X^2}
\bigg{)}\bigg{]}X^{\prime}_{\mu}\bigg{\}}\,,}\eqno(\new)$$
where $\delta\,(<<\Delta)$ is the width of the string core and corresponds to
the
ultra-violet renormalisation scale. These expressions  for the self and
radiation force densities are extremely complicated, however, our
confidence that these are the correct expression is strengthened  since  they
non-trivially respect the conformal gauge conditions, that is, ${\cal F}^{\rm
self}\cdot\dot X=0$, ${\cal F}^{\rm self}\cdot X^{\prime}=0$,
${\cal F}^{\rm rad}\cdot\dot
X=0$ and ${\cal F}^{\rm rad}\cdot X^{\prime}=0$.

The self-field is a multiple of the left hand side of the equations of
motion and facilitates the well-known renormalisation of the string tension,
in a way exactly analogous to the mass of a point electron. The equations of
motion in this case are
\eqnam{\confreneom} $$ \mu(\Delta)\bigg{[}\ddot
X_{\mu}-X^{\prime\prime}_{\mu}\bigg{]}={\cal F}^{\rm rad}_{\mu}\,,\eqno(\new)$$
where $\mu(\Delta)=\mu_0+2\pi\fa^2\log(\Delta/\delta)$
is the renormalised string tension

For general string trajectories, similar to those discussed in
ref.\refto{BSa}, some of the terms in (\backforcetwo) can be shown to be
sub-dominant. In particular, if the string solution is specified by the
relative amplitude ${\cal E}$ and its wavelength $L$, then one finds that
${\bf X}\sim {\cal O}({\cal E} L)$ and each subsequent derivative requires a
division by $L$. Most of the higher order terms in ${\cal E}$ can be dropped,
though it is necessary to keep two of the higher order terms to maintain the
gauge conditions. One then finds that it is possible to approximate $H^{\rm
rad}_{\mu\alpha\beta}$ and  ${\cal F}_{\mu}^{\rm rad}$ by
\eqnam{\confapproxforce}   $$  \eqalign{H^{\rm
rad}_{\mu\alpha\beta}&\approx-{2\fa\Delta\over 3\dot
X^4}X^{\tdot}_{[\mu}V_{\alpha\beta]}\,,\cr  {\cal F}^{\rm
rad}_{\mu}&\approx{4\pi\fa^2\Delta\over 3}\bigg{[}X^{\tdot\mu}-
{\bigg{(}{\dot X\cdot X^{\tdot}\over 1-\dot X^2}\bigg{)}}\dot X_{\mu}+
{\bigg{(}{X^{\prime}\cdot X^{\tdot}\over 1-\dot
X^2}\bigg{)}X_{\mu}^{\prime}\bigg{]}} \,.}\eqno(\new)  $$

\subhead{2.6 Generalization to the temporal transverse gauge}

\noindent For flat-space string dynamics, the conformal string gauge is usually
employed. However, when considering problems in which the string energy decays,
it more convenient to use the temporal transverse gauge in which $X^0=t=\tau$
and  $\dot {\bf X}.{\bf X}^{\prime}=0$ with $X^\mu=(t,{\bf X})$. In this gauge,
the equations of motion for the string (\modeom) are
\eqnam{\tempeom}  $$ \mu_0\bigg{(}\ddot{\bf X}-{1\over \epsilon}\bigg{(}{{\bf
X}^{\prime}\over\epsilon}\bigg{)}^{\prime}\bigg{)} = {\bf f}\,,\quad
\mu_0\dot\epsilon = f^0\,,\eqno(\new) $$
where $\epsilon^2 = \dot {\bf X}^2/(1-{\bf X}^{\prime 2})$ (not to be
confused with the relative amplitude ${\cal E}$) and
${\cal F}^{\mu}=(f^0,\epsilon{\bf f}+f^0\dot{\bf X})$.  Radiative damping will
naturally be incorporated in the decay of the coordinate energy density
$\epsilon$, rather than in the non-intiutive time redefinitions of the
conformal
gauge. The string energy and
momentum per unit length in the temporal transverse gauge are then given by
\eqnam{\energyandmom} $$E={\mu_0\over L}\int_0^L d\sigma\epsilon\,,\quad {\bf
p}={\mu_0\over L}\int_0^L d\sigma\epsilon\dot{\bf X}\,.\eqno(\new)$$

The renormalisation procedure in this gauge is similar to that for the
conformal
gauge, but with the added complication that $\dot X^2+X^{\prime 2}\ne 0$. After
a detailed set of manipulations one can deduce that
\eqnam{\renfields} $$  \eqalign{
H_{\mu\alpha\beta}^{\rm self}=&{\fa\over 2\dot X^4}\bigg{[}{1\over
\epsilon}\ddot
X_{[\mu}V_{\alpha\beta]} -
{1\over\epsilon^3}X^{\prime\prime}_{[\mu}V_{\alpha\beta]}\bigg{]}\log(\Delta/\delta)+{\cal
O} (\Delta^2)\,,\cr H_{\mu\alpha\beta}^{\rm rad}=&{\fa\over 2\dot
X^4}\bigg{[}-{4\over 3}X^{\tdot}_{[\mu}V_{\alpha\beta]}-{1\over 2}\dot
X_{[\mu}\dot V_{\alpha\beta]}+{3\dot\epsilon\over
2\epsilon}X^{\prime\prime}_{[\mu}V_{\alpha\beta]}\cr&+\bigg{(} {2\dot X.\ddot
X\over \dot X^2} - {\dot\epsilon\over 2\epsilon}\bigg{)}\ddot
X_{[\mu}V_{\alpha\beta]}\bigg{]}\Delta+{\cal O} (\Delta^2)\,,\cr}\eqno(\new)
$$   where
$A_{[\mu\alpha\beta]}=A_{\mu\alpha\beta}+A_{\beta\mu\alpha}+A_{\alpha\beta\mu}$.
Note that, once again, the self-field has no order $\Delta$ term. Ignoring
terms
of order $\Delta^2$, we can then deduce expressions for the self-force and the
radiation backreaction force,
\eqnam{\tempforce}
$$
\eqalign{{\bf f}^{\rm self} =& -2\pi\fa^2\log(\Delta/\delta)\bigg{(}\ddot{\bf
X}-{1\over\epsilon}\bigg{(}{{\bf
X}^{\prime}\over\epsilon}\bigg{)}^{\prime}\bigg{)}\,,\cr
{\bf f}^{\rm rad} =
& \pi\fa^2\Delta\bigg{\{}{4\over 3}\epsilon{\bf X}^{\tdot}
+\bigg{[}2\epsilon\bigg{(}{\dot{\bf X}\cdot\ddot{\bf X}\over 1-\dot{\bf X}^2}
\bigg{)} + 3\dot\epsilon\bigg{]}\ddot{\bf X} - {2\over\epsilon}\bigg{(} {{\bf
X}^{\prime}\cdot\ddot{\bf X}\over 1-\dot{\bf X}^2}\bigg{)}{\dot{\bf
X}}^{\prime}
-{3\dot\epsilon\over\epsilon^2}{\bf X}^{\prime\prime}\cr
+& \bigg{[}-{4\over
3\epsilon}\bigg{(}{{\bf X}^{\prime}\cdot {\bf X}^{\tdot}\over 1-\dot{\bf
X}^2}\bigg{)} - {4\over\epsilon}{(\dot{\bf X}\cdot\ddot{\bf X} )({\bf
X}^{\prime}\cdot\ddot{\bf X})\over (1-\dot{\bf X}^2)^2} - {\dot\epsilon
\over\epsilon^2}\bigg{(}{{\bf X}^{\prime}\cdot\ddot{\bf X}\over 1 -\dot{\bf
X}^2}
\bigg{)}+{3\dot\epsilon\over\epsilon^4}\bigg{(}{\dot{\bf X}\cdot{\bf X}^{\prime
\prime}\over 1 -\dot{\bf X}^2}\bigg{)}\bigg{]}{\bf X}^{\prime}\bigg{\}} \,,\cr
f^{0,\rm self}=& -2\pi\fa^2\log(\Delta/\delta)\dot\epsilon\,,\cr
 f^{0,\rm
rad}=& \pi\fa^2\Delta\bigg{\{}{4\over 3}\epsilon^2\bigg{(}{\dot{\bf X}\cdot
{\bf X}^{\tdot}\over 1-\dot {\bf X}^2}\bigg{)}+2\bigg{(}{{\bf
X}^{\prime}\cdot\ddot
{\bf X}\over 1-\dot {\bf X}^2}\bigg{)}^2+2\epsilon^2\bigg{(}{\dot{\bf
X}\cdot\ddot{\bf X} \over 1 -\dot{\bf X}^2}\bigg{)}^2 \cr
&+3\epsilon\dot\epsilon\bigg{(}{\dot{\bf X} \cdot\ddot{\bf X}\over 1-\dot{\bf
X}^2}\bigg{)} -{3\dot\epsilon\over\epsilon} \bigg{(}{\dot{\bf X}\cdot{\bf
X}^{\prime\prime}\over 1 -\dot{\bf X}^2}\bigg{)}\bigg{\}} \,.}\eqno(\new)
$$
As in the conformal gauge, the expressions for ${\bf f}^{\rm self}$
and $f^{0,\rm self}$, facilitate the well-known renormalisation of the
equations
of motion (\tempeom) and coordinate energy density, \eqnam{\reneom}  $$
\mu(\Delta)\bigg{(}\ddot{\bf X}-{1\over \epsilon}\bigg{(}{{\bf
X}^{\prime}\over\epsilon}\bigg{)}^{\prime}\bigg{)} = {\bf f}^{\rm rad}\,,\quad
\mu(\Delta)\dot\epsilon = f^{0,\rm rad}\,,\eqno(\new) $$  where the expressions
for ${\bf f}^{\rm rad}$  and $f^{0,\rm rad}$ represent the finite radiation
backreaction force. The renormalised versions of (\energyandmom) are
\eqnam{\actenergyandmom}  $$E={\mu(\Delta)\over L}\int_0^L
d\sigma\epsilon\,,\quad{\bf p}={\mu(\Delta)\over L}\int_0^L
d\sigma\epsilon\dot{\bf X}\,.\eqno(\new)$$ Differentiating (\actenergyandmom),
gives the power and force due to radiation backreaction \eqnam{\powerandforce}
$$\eqalign{\dot E&={\mu(\Delta)\over L}\int_0^L d\sigma\dot\epsilon={1\over
L}\int_0^L d\sigma f^{0,\rm rad}\,,\cr \dot {\bf p}& = {\mu(\Delta)\over
L}\int_0^L d\sigma\bigg{[}\epsilon\ddot{\bf X}+\dot\epsilon\dot{\bf
X}\bigg{]}={1\over L}\int_0^L d\sigma\bigg{[}\epsilon{\bf f}^{\rm rad}+f^{0,\rm
rad}\dot {\bf X}\bigg{]}\,.}\eqno(\new)$$

Again, some of the terms in
(\renfields) and (\tempforce) can be shown to be sub-dominant, as for the
conformal gauge. In particular it is possible to approximate $H^{\rm
rad}_{\mu\alpha\beta}$, ${\bf f}^{\rm rad}$ and $f^{0,\rm rad}$ by
\eqnam{\approxforce}   $$
\eqalign{H^{\rm
rad}_{\mu\alpha\beta}&\approx-{2\fa\Delta\over 3\dot
X^4}X^{\tdot}_{[\mu}V_{\alpha\beta]}\,,\cr
{\bf f}^{\rm
rad}&\approx{4\pi\fa^2\Delta\over 3}\bigg{[}\epsilon{\bf X}^{\tdot}- {1\over
\epsilon} \bigg{(}{{\bf X}^{\prime}\cdot{\bf X}^{\tdot}\over 1-\dot {\bf
X}^2}\bigg{)}{\bf X}^{\prime}\bigg{]}\,,\cr
f^{0,\rm rad}&\approx{4\pi\fa^2\Delta\over
3}\bigg{[}{\epsilon^2\dot {\bf X}\cdot{\bf X}^{\tdot}\over 1-\dot {\bf
X}^2}\bigg{]}
\,.}\eqno(\new)
$$
Substituting the expressions for $f^{0,\rm rad}$ into the power expression
(\powerandforce) yields
\eqnam{\poweract}
$${dP\over dl}=-\dot E= -{4\pi\fa^2\Delta\over 3L}\int_0^L
d\sigma{\epsilon^2\dot{\bf X}.{\bf X}^{\tdot}\over 1-\dot{\bf
X}^2}\,.\eqno(\new)$$

\subhead{2.7 Eliminating `runaway' solutions}

\noindent This simplified form of the equations of motion using (\approxforce)
still has serious shortcomings because of the presence of the ${\bf X}^{\tdot}$
term.  The equations  have unphysical, exponentially growing or `runaway'
solutions which will, for example, plague any potential numerical
applications.   Furthermore, one would be required to store information at
three different timesteps, fundamentally changing the nature of a numerical
algorithm. It appears, however, that both these problems can be circumvented
by resubstituting the equations of motion, that is, we make the approximations
$\ddot {\bf X} \approx \epsilon^{-1}({\bf X}'/\epsilon)'$ and $ {\bf X}^{\tdot}
\approx \epsilon^{-1}(\dot {\bf X}'/\epsilon)'$ in (\approxforce) (note we
have used the unperturbated equations with $\dot \epsilon \approx 0$).  The
equations of motion then acquire an analogue of a viscosity term for which
there
are only damped solutions. After performing this resubstitution one finds that
the approximate force (\approxforce) becomes  \eqnam{\newapproxforce}
$$
\eqalign{{\bf f}^{\rm
rad}&\approx{4\pi\fa^2\Delta\over 3}\bigg{[}
\left({\dot{\bf
X}^{\prime}\over \epsilon}\right)^{\prime}- {1\over \epsilon^2} \bigg{(}{{\bf
X}^{\prime}\cdot(\dot {\bf X}'/\epsilon)'\over 1-\dot {\bf X}^2}\bigg{)}{\bf
X}^{\prime}\bigg{]}\,,\cr
f^{0,\rm rad}&\approx{4\pi\fa^2\Delta\epsilon\over
3}\bigg{[}{\dot {\bf X}\cdot(\dot {\bf X}'/\epsilon)'\over 1-\dot {\bf
X}^2}\bigg{]} \,.}\eqno(\new)  $$

\figure{runaway.eps}{2.2in}{0.75in}{3}{The relative positions of the curves
$y=f(m)$ (solid line) and $y=g(m)$ (dotted line) for typical values of the
parameters $\Omega$ and $\alpha$. Notice that the real positive solutions of
$f(m)$=0---corresponding to the exponentially growing solution of the equations
of motion---is not a solution of $g(m)=0$.}

The reason for the suppression of the exponentially growing solution becomes
apparent if we consider simplified one-dimensional equations,  \eqnam{\simpeqn}
$$ \ddot X - X^{\prime\prime} = \alpha X^{\tdot}\,,\quad\longrightarrow\quad
\ddot X -  X^{\prime\prime} \approx \alpha \dot X^{\prime\prime} \,,\eqno(\new)
$$   where we have performed the resubstitution assuming that $\alpha$ is
small.
We now take an approximately periodic solution,    $X^{\prime\prime}\approx
-\Omega^2X$, and we substitute the ansatz $X\sim e^{mt}$. The solutions for
(\simpeqn) are given respectively by the roots of the following polynomials
in $m$,   \eqnam{\poly} $$f(m)=\alpha
m^3-m^2-\Omega^2\,,\quad\longrightarrow\quad
g(m)=-m^2-\alpha\Omega^2m-\Omega^2\,.\eqno(\new)$$
 If we rewrite $f(m)=(m^2+Am+\Omega^2+B)(\alpha m-C)$, then we see  that
$A=\alpha\Omega^2+{\cal O}(\alpha^2)$, $B={\cal O}(\alpha^2)$ and $C=1+{\cal
O}(\alpha^2)$. If we ignore terms ${\cal O}(\alpha^2)$, then the solutions of
$g(m)=0$ are approximately solutions of $f(m)=0$. However, the real positive
solution of $f(m)=0$, corresponding to the exponentially growing solution of
the
equations of motion, is not a solution of $g(m)=0$. Fig. 3 shows the
relative positions of the curves $y=f(m)$ and $y=g(m)$ for typical values of
the parameters $\Omega$ and $\alpha$.

\subhead{2.8 Understanding the `local backreaction approximation'}

\figure{local_back.eps}{2.2in}{-0.35in}{4}{A schematic of the contributions to
the radiation backreaction force for a perturbed long string configuration. (a)
The force is calculated by integrating along $\sigma$, that is, summing the
contributions of all string segments on the backward light cone. (b) The
expected
appearance of the actual radiation force contributions; this has a local
maximum
and a fairly rapid fall-off with cancellations. Here, $L$ is the typical
wavelength of perturbations on the string (or the string curvature radius) and
the area under the curve is the total magnitude of the  force. (c) The
radiation
force is estimated in the local backreaction approximation using the local
magnitude of the force and an effective width $\Delta$ for which the area under
the two curves is equal.  For this approximation to be valid, $\Delta$ must be
less than the string curvature radius.}

\noindent The local backreaction approximation effectively reduces the problem
of
calculating the backreaction force from the string to the equivalent problem
for
the electron. The assumptions underlying (\backforcetwo) and (\reneom)  are
that
the dominant contributions to the integrals (\bsoln), (\lwpot) and
(\lwpotselfrad) come from string segments close to the point under
consideration. In fig.~4 we have schematically illustrated the construction of
the radiation backreaction force in the local backreaction approximation.  To
calculate this force at the time $t$ at a particular point on the string (say
$\sigma=0$), we must sum over the retarded time contributions
from all other string segments.  For definiteness, let us suppose we are
considering a long straight string with perturbations of typical wavelength $L$
(comparable to the string curvature radius $R$).  As we integrate along the
backward light cone of fig.~4a, we can expect force contributions to take a
form
appearing something like fig.~4b.  The precise rate of the fall-off
of this force density away from $\sigma=0$ is unknown, but finiteness certainly
implies that it is faster than $1/\sigma$.  Moreover, regions of the string
beyond the curvature radius $R$ will give negative, as well as positive,
contributions and the resulting net cancellations should ensure rapid
convergence of the integrated force at large distances.

The total area under the curve in fig.~4b represents the exact magnitude of the
radiation backreaction force.  The `local backreaction approximation' to this
force is illustrated in fig.~4c.  We calculate the actual magnitude of the
force at the point in question and we assume that the contribution from
neighbouring segments falls away rapidly beyond an effective width $\Delta$.
We then normalize $\Delta$ to ensure that the area under the curve (c) is equal
to that under curve (b).  In \S3 we shall discuss the procedure for achieving
this by comparing with analytic and numerical radiation calculations.

Of course, we do not expect our local force to evolve every string
trajectory completely accurately.  We are, after all, assuming a uniform
$\Delta$, though it is possible to improve this first approximation.
Furthermore, the force given by (\approxforce), which we shall use in
the numerical simulations of \S3, breaks worldsheet covariance and so we should
anticipate difficulties describing some special `null'
string trajectories. This is because we have taken time as a preferred
direction
in the derivation of the Lienard-Wiechart potentials. Effectively, we calculate
the backreaction force by summing up all contributions from points on the
string
inside the region $D$ given by
\eqnam{\sigtau}
$$
D=\bigg{\{}(\sigma,\tau)\hbox{
s.t.
}-\Delta/2<\sigma<\Delta/2\,,\,-\Delta/2<\tau<\Delta/2\bigg{\}}\,,\eqno(\new)
$$
which corresponds to the shaded diamond in fig.~3a.
However, there exists a family of elongated rectangular regions of equal area
related to $D$ by Lorentz boosts. If the string trajectory has a typical
wavelength (or periodicity) then it seems likely that the contributions from
the
two regions will be similar and the inaccuracies in the force should
cancel out.  We anticipate, therefore, that errors in the local backreaction
approximation should be small for generic long string trajectories and for
closed loops where this `pseudo-periodicity' is likely to be evident.

In essence the validity of the local
backrection approximation hinges on whether the following integral is small:
\eqnam{\lwpotderivs}
$$
\partial_{\mu}B_{\alpha\beta}(x)= -{\fa\over
2}\int_{|\sigma-\bar\sigma|>\Delta/2}\,  d\bar\sigma  {1\over \Delta\cdot\dot
X}{\partial\over \partial\bar\tau}\bigg{(}{\Delta_{\mu}V_{\alpha\beta}\over
|\Delta\cdot\dot X|}\bigg{)}\bigg{|}_{\bar\tau=\tau_R}\,.\eqno(\new)
$$
To summarize, we believe that (\lwpotderivs) can be neglected for
strings in a realistic network because the natural long distance force fall-off
will be augmented by strong cancellations from a random superposition of
distant
modes.  In any case, at the very least, this approximation should work in an
some `averaged' sense.

\subhead{2.9 Analytic models for radiative decay}

\noindent We can begin to develop confidence in the veracity of the local
backreaction approximation by demonstrating that it predicts the correct
scale-dependence of the overall radiation power from closed loops and long
string trajectories.
By analogy with the simple model for
backreaction of ref.\refto{BSa}, we shall deduce expressions for the evolution
of the invariant string length $L$ for loops and the relative amplitude
${\cal E}$ for perturbed long strings, using some of the solutions we
presented in \S2.2.

\medskip
\noindent{\it (i) Closed loop solutions}
\medskip

\noindent For closed loops of invariant length L,  we can estimate that
$\dot{\bf
X}\sim{\cal O}(1)$ and ${\bf X}^{\tdot}\sim {\cal O}(L^{-2})$ (the actual
average over one period is $\langle |\dot{\bf X}|\rangle=1/\sqrt2$).
If we take $\Delta\sim
L$ in our local approximation, then the power per unit length (\poweract) is
proportional to $L^{-1}$, which immediately recovers the well known result that
the power loss from a loop is independent of its size $L$\refto{VV}. In
general,
one can write the radiation power as \eqnam{\defgamma}
$$P=\Gamma_a\fa^2=\kappa\mu\,,\eqno(\new)$$ where $\Gamma_a$ is some factor
dependent on the particular loop trajectory, but not its size and $\kappa$ is
the radiation backreaction scale, assumed to be independent of
time\footnote{\dag}{Problems involving global strings become intractable if the
logarithmic time dependence of the string tension is  included. In cosmological
problems one finds that the logarithm changes by  only about one order of
magnitude over the enormous timescale between string formation and the present
day. We will make this assumption in all the following
calculations.\vskip 0pt}. The radiation damping can be modelled by considering
the following equations, \eqnam{\loopmodel} $$\eqalign{E&=\mu\int_0^L
d\sigma\epsilon=\mu L\,,\cr P&=-{dE\over
dt}=\Gamma_a\fa^2=\kappa\mu\,.}\eqno(\new)$$ Integrating these equations one
obtains, \eqnam{\loopform}
$$L=L_0-\kappa (t-t_0)\,.\eqno(\new)$$

\medskip
\noindent{\it (ii) Long string solutions}
\medskip

\noindent The generic result for a long string solution, parametrized by  its
wavelength $L$ and relative amplitude ${\cal E}$, is $\dot {\bf X}\sim {\cal
O}({\cal E})$ and ${\bf X}^{\tdot}\sim {\cal O}({\cal E} L^{-2})$.
If $\Delta\sim L$, then the power per unit length in (\poweract) is
\eqnam{\longpower}
$${dP\over dl} ={\beta{\cal E}^2\over L}\,,\eqno(\new)$$
where $\beta\sim\fa^2$ quantifies the overall strength of the radiation.
By analogy to the closed loops the radiation damping can be
modelled by, $$\eqalign{E&={\mu\over L}\int_0^L d\sigma\epsilon=\mu
+\alpha\mu{\cal E}^2\,,\cr {dP\over dl}&=-{dE\over
dt}={\beta{\cal E}^2\over L}\,,}\eqno(\new)$$
where
$\alpha$ is the (order unity) solution-dependent coefficent of ${\cal E}^2$ in
the power series expansion of
\eqnam{\alphadef}
$$\bigg{[}{1\over L}\int_0^L{dX^3\over
d\sigma}d\sigma\bigg{]}^{-1}\,.\eqno(\new)$$
The  power loss (\longpower)
will lead to an exponential decay of the amplitude and oscillation energy per
unit length $E$,
\eqnam{\newmodel}
$$
{\cal E}={\cal E}_0 \exp\bigg{(}-{\beta t\over 2\alpha\mu
L}\bigg{)}\,,\quad E = \mu+\alpha\mu{\cal E}_0^2 \exp\bigg{(}-{\beta
t\over\alpha\mu L}\bigg{)}\,.\eqno(\new)
$$

Exponential decay has already been shown to be generic\refto{BSa} for a
realistic
situation  where the left- and right-moving amplitudes are not
precisely the same as in (\leftrightsolndef). However, for some of the
periodic solutions of \S2.2 such as the `standing wave'
solutions (\helixsolndef) and (\cosinesolndef), the amplitude fall-off was
shown
analytically to be a power law, a fact also confirmed in numerical field
theory simulations\refto{BSa}. How can we reconcile this
apparent contradiction? The answer lies in noting that the local
backreaction approximation only applies to string configurations in which
long-range field correlations are suppressed beyond the string curvature
radius.
Thus when we compare the effects of the radiation backreaction force with
numerical field theory simulations, we must ensure that global correlations are
suppressed, eliminating artificial situations with large-scale periodic
coherence.

\sectbegin{3}{Numerical Comparisons}

\subhead{3.1 Numerical methods}

\noindent In order to solve the modified equations of motion (\reneom) and
(\newapproxforce) numerically, one must recast the simplified resubstituted
equations of motion into a first-order form accessible to numerical solution.
Defining ${\vec\alpha}={\bf X}^{\prime}-\epsilon\dot {\bf X}$ and
${\vec\beta}={\bf X}^{\prime}+\epsilon\dot {\bf X}$, the equations of motion
can
be rewritten as \eqnam{\numericaleom}  $$
\eqalign{\mu(\Delta)\bigg{[}\dot{\vec\alpha}+\bigg{(}{{\vec\alpha}
\over\epsilon}\bigg{)}^{\prime}\bigg{]}&=-{1\over 2}\bigg{(}\epsilon{\bf
f}^{\rm rad}+f^{0,\rm rad}\dot {\bf X}\bigg{)}\,,\cr
\mu(\Delta)\bigg{[}\dot{\vec\beta}-\bigg{(}{{\vec\beta}\over\epsilon}
\bigg{)}^{\prime}\bigg{]}&={1\over 2}\bigg{(}\epsilon{\bf f}^{\rm
rad}+f^{0,\rm rad}\dot {\bf X}\bigg{)}\,,\cr \mu(\Delta)\dot\epsilon =
f^{0,\rm rad}\,,~~&\qquad \dot{\bf X} = {1\over
2\epsilon}\bigg{(}\vec\beta-\vec\alpha\bigg{)}\,.}\eqno(\new)
$$

Using the above, we were able to evolve string trajectories by modifying
a total variation non-increasing (TVNI) algorithm\refto{sod,ASa} which has
already been well-tested for string network evolution in an expanding
universe. This method relies on the fact that the first order equations of
motion (\numericaleom) are in conservative form, if the backreaction force
is zero. Artificial
compression methods are used to prevent the numerical dissipation of kinks.
Typically, the algorithm maintains the perpendicularity condition $\dot{\bf
X}\cdot{\bf X}^{\prime}=0$ and conserves energy
to within a few percent over many timesteps. It should be noted that the
addition of a small backreaction forcing term does not seem to affect the
stability of the numerical scheme.  However, if the backreaction force
becomes larger than the tension force, then the equations of motion become
qualitatively different behaving like a diffusion equation rather than a
hyperbolic wave equation. The characteristic Courant condition for a
diffusion equation is very much more restrictive and so
stability problems will emerge in this regime. One can address this numerical
problem by artifically preventing the force from becoming too large, that is,
the usual procedure of `force softening'.  There is a further technical
numerical
problem because the coordinate energy $\epsilon$ decays more rapidly at certain
points (for example, at the cusps of the Kibble--Turok loops).  Eventually,
this
imposes an unacceptably small timestep on the simulation because we always
require $\Delta t < \epsilon \Delta\sigma$ everywhere.  However, this problem
can be solved by a number of approaches, including multiple time-stepping in
small $\epsilon$ regions, by reparametrising the string to redistribute
$\epsilon$ more evenly, or by eliminating such regions through `point-joining'
techniques.

In all the numerical simulations using the radiation backreaction force in
this paper, we have employed the constant damping coefficent given by
\eqnam{\dampcoeff}
$${4\pi\fa^2\Delta\over 3\mu(\Delta)}\approx 0.001L\,.\eqno(\new)$$ This choice
of damping coefficient corresponds to the cosmologically interesting
paremeter range  $\mu(\Delta)\sim 100\fa^2$ with our numerically determined
normalization $\Delta\sim 0.1L$ which we shall discuss in the next section.
When we compare the results of these Nambu string simulations with those using
the underlying field theory for which $\mu(\Delta)\sim 5\fa^2$, we have
had to perform a single global rescaling of the time axis in order to take into
account the different radiation strengths.

As well as this one-dimensional effective model, we have developed
sophisticated numerical algorithms to dynamically simulate string
configurations in the Goldstone model (\goldaction)\refto{BSa,Sd,BSd}.
We  discretize space on a three-dimensional grid with dimensions $N_1,N_2, N_3$
in the $x,\,y,\,z$ directions respectively, solving the rescaled ($f_a\to 1,~
\lambda\to 2$) Euler-Lagrange equation, \eqnam{\elrescale}
$${\partial^2\Phi\over\partial t^2} - {\partial^2\Phi\over\partial x^2} -
{\partial^2\Phi\over\partial y^2} - {\partial^2\Phi\over\partial z^2}
+\Phi(\bar\Phi \Phi-1) = 0\,.\eqno(\new)$$ We employ a second-order leapfrog
algorithm for the time derivative  and fourth-order finite difference
approximations for the spatial derivatives. In problems  where radiation is
incident on the boundaries, its sensible to use absorbing boundary
conditions\refto{BSa,BSd}. A second order wave equation which annihilates the
reflected wave at the $x=0$ boundary is
\eqnam{\boundarycond}
$$
{\partial\over\partial t}{\partial\Phi\over\partial x} - {\partial^2\Phi\over
\partial^2t} + {1\over 2}\left({\partial^2\Phi\over\partial^2y} +
{\partial^2\Phi \over\partial^2z}\right)=0\,.\eqno(\new)
$$
The efficacy of these methods is discussed in some detail in ref.\refto{BSa}.

We use a cylindrically symmetric string ansatz to create initial
conditions for both long string  and loop solutions as in ref.\refto{BSa}.
However, a naive application of this ansatz artificially creates
long-range correlations  which do not conform with the assumptions underlying
the `local backreaction approximation'. As we have emphasised, general
configurations that occur in realistic string networks will not have field
correlations beyond the average curvature radius of the string because of
reconnection processes and causality constraints. Consequently, we have
modified our ansatz for long string configurations by numerically
suppressing the initial perturbations with the gaussian,
\eqnam{\supression}
$$
e^{-(|{\bf x}-{\bf
x}_{\rm s}|/R)^2}\,,\eqno(\new)
$$
where $R$ is the string curvature radius, ${\bf x}$ is the position in question
and ${\bf x_{\rm s}}$ is the nearest long string segment.  At large distances
$r\hbox{$>\!>$} R$, therefore, the string fields will approach those for a
straight string, as we would expect in a general physical context for random
small-scale structure.

\figure{supunsup.eps}{4.0in}{0.05in}{5}{A comparison of the decay amplitude
for a sinusoidal solution in full field theory (a) without suppression and (b)
with suppression of the field at the curvature radius.
Notice that the initial decay rate of the suppressed configuration is much
faster (exponential) than that for the unsuppressed configuration (power law)
due to the long range correlations of the latter.}

To make this distinction plain, fig.~5 illustrates the effect of large distance
correlation suppression on the decay of a periodic sinusoidal solution.  The
suppressed case (fig.~5b) can be seen initially to decay more rapidly than the
configuration in which perturbations in the fields are correlated out to the
simulation boundary (fig.~5a); the former is exponential decay, while the
latter
is power law.   However, given the periodic boundary conditions, this
difference
does not persist indefinitely because the suppressed configuration will
causally
relax to an unsuppressed one, as the long range fields become correlated on
larger and larger scales. Given this limitation imposed by the
numerical grid size, we can only expect to normalize the local
backreaction approximation using relatively short simulations (or by using
non-periodic configurations).

\subhead{3.2 Long string configurations}

\figure{loglin.eps}{4.5in}{0.0in}{6}{Log-linear plots for the decay of
amplitude for (a) a sinusoidal solution, (b) a pure left moving helicoidal
perturbation, and (c) a helicoidal perturbation with unequal left- and right-
moving amplitudes. The straight line typifies the exponential decay.}

\figure{decay.eps}{4.0in}{0.45in}{7}{Decay of ${\cal E}$ using the
radiative backreaction force (dotted line) and numerical field
theory simulations (solid line) for (a) a sinusoidal perturbation, (b) a
helicoidal perturbation with unequal left- and right-moving amplitudes, and (c)
a pure left moving helicoidal perturbation. Note the excellent quantitative
agreement for all three cases.}

\noindent We have extensively tested the `local backreaction approximation',
using the modified Nambu equations of motion (\numericaleom) and the
approximate
force derived from (\approxforce) by directly comparing with field theory
simulations of  radiating strings in the Goldstone model.

Using our modified Nambu string simulations, we find that exponential decay is
generic for all the long string configurations discussed in \S2. This
decay is illustrated in fig.~6 for the sinusoidal solution (\cosinesolndef),
the
pure right-moving helicoidal solution (\rightsolndef) and the helicoidal
solution
with unequal left-  and right-moving amplitudes (\leftrightsolndef).
Fig.~7 illustrates the  excellent quantitative agreement with the full
field theory simulations by direct comparison with the same three long
string solutions.  Note that these curves have not been matched separately; the
same backreaction damping coefficient applies for each and there has been only
a single global rescaling. The agreement persists for the longest time
for the (generic) unequal left- and right-moving configuration
(\leftrightsolndef)
because exponential decay is predicted in this case even after field
correlations
have relaxed at large distances.  By comparing with the simple backreaction
model for exponential decay (\newmodel),  one can use the numerical field
theory
results to normalize $\Delta$, that is, we  estimate
\eqnam{\Deltalongstring}
$$
\Delta\approx
(0.1\pm 0.02)L\,\eqno(\new)
$$
for the long string solutions investigated.
This is a result for which there are considerable uncertainties at this stage,
mainly because of the imprecision inherent in our small-scale field theory
simulations.
We had anticipated that $\Delta$ should be normalized to a distance
near the string radius of curvature $R$, which for a sinusoidal perturbation is
$R\sim L/4$.  The fact that $\Delta$ is smaller than $R$ validates the
linearized expansion on which the `local backreaction approximation' is based.
The normalisation of the value of $\Delta$ above can become ambiguous in
certain physical contexts, such as a solution with a number of different
Fourier modes. In this case, we must make a further approximation by
normalising to the lengthscale which is radiatively dominant.

We have also applied these numerical approaches to study the kink
solution (\kinksolndef) of \S2. Fig.~8 compares the evolution of a sharp kink
in
both the local backreaction approximation and in a field theory simulation.
Backreaction leads to a substantial rounding of the kink, in agreement with the
intuitive picture described in refs.\refto{Hina,BSa}. The results are almost
indistinguishable except for the computational advantages of the former which,
in this case, saved a factor of $10^2$ in cpu time and $10^4$ in allocated
memory.

We have also performed spectral analysis of the modes on the string using
techniques similar to those used in three dimensions in ref.\refto{BSa}. The
kink itself can be written as an infinite series of odd Fourier modes, while
the anticipated endpoint, a sinusoidal solution, is just a single Fourier mode.
Fig.~9 illustrates the mode decomposition of the kink solution initially and
then at late times.  Radiation backreaction causes decay in all modes, but the
higher harmonics are clearly damped much more strongly, leading to the kink
`rounding'.  These spectra can be compared to the kink radiation fields shown
in
ref.\refto{BSa} which demonstrate the same trend.

\vdoublefigure{nam.eps}{field.eps}{1.5in}{0.8in}{1.2in}{1.5in}{8} {Decay of a
kink perturbation $({\cal E}_0$=$0.9)$ using (a) the radiation
backreaction force and (b) numerical field theory. Notice the visible
rounding of the kink in both cases.}

\figure{spec_kink.eps}{4.0in}{0.6in}{9}{The time
evolution of the Fourier modes of an initial kink configuration (a). Note the
damping of higher modes after 16 oscillations in (b).}

\subhead{3.3 Closed loop solutions}

\noindent We have also applied the local backreaction approximation to the
study of loop solutions, such as the Kibble-Turok loops described in \S2.2. In
this case, $\Delta\sim L$ is not independent of time because the loop shrinks
as
it decays\footnote{*}{For the long string solutions, periodic boundary
conditions
forced the solution to have a fixed time-independent wavelength $\lambda \sim
L$.}. This problem can be circumvented in the case of loops by choosing the
cut-off scale $\Delta$ equal to a constant multiple of the total invariant
string length,  \eqnam{\newdelta} $$ L = \int_0^{2\pi}
d\sigma\,\epsilon\,,\eqno(\new)$$ which is easily calculable within the
evolution algorithm described earlier.

\figure{sqpict2.eps}{1.3in}{-0.1in}{10}{The time
evolution of a kinky loop solution shown initially and after 5 and 10
oscillations.  Note the decrease in loop size and discernible kink `rounding'.}

The damped evolution of the special kinky loop solution (\akink) is shown after
several oscillations in fig.~10.  As the loop shrinks in size, there is
discernible `rounding' due to the radiative damping, though it is less
pronounced
than in the long string kink decay.  Unfortunately, evolution for this and
other
loops could not be continued indefinitely because a numerical solution to the
Courant violation problem in small $\epsilon$ regions has yet to be implemented
(refer to \S3.1).  However, the observed `rounding' is at least qualitatively
in
agreement with a previous attempt to study gravitational backreaction in
ref.\refto{QS}.  In this non-local approach, all the retarded time radiation
contributions were accumulated for an unperturbed loop trajectory and then
these
`corrections' were applied at the end of each oscillation period.  Unlike the
local backreaction approximation, there is little prospect of such `exact'
approaches being implemented in network simulations because the ${\cal O}(N^2)$
algorithms require a supercomputer to evolve a single loop.  Nevertheless, we
anticipate future quantitative comparisons with such methods to determine the
accuracy of our approach.

\figure{straight.eps}{3.5in}{0in}{11}{The effect of radiative damping on loop
energy for $\alpha=0$ Kibble--Turok loops with $\phi = \pi/12$ (solid
line), $\pi/3,$ (dotted line) and $5\pi/12$ (dashed line). Note the expected
linear decay of the loop length.}

The evolution of the energy of some Kibble--Turok loops is illustrated in
fig.~11.  One can readily observe the linear decay of these solutions, as
expected from our simple backreaction model (\loopform).  Notice, however, the
oscillatory nature of the decay due to stronger radiation when the loop
trajectories becomes more convoluted and when cusps appear.  It is interesting
to
note that a preliminary analysis indicates that, while cusp velocities are
curtailed by backreaction, their periodic reappearance in these particular
solutions is not actually prevented.

The overall decay rate in fig.~11 is parameter dependent; the slope yields the
backreaction scale $\kappa$ (or $\Gamma_a$) in (\defgamma) which is appropriate
for the particular loop trajectory.  Fig.~12 illustrates the $\phi$-dependence
of $\kappa$ for the $\alpha=0$ loop solutions (\kibbleturoksolndef).  This is
qualitatively similar to analytic estimates of the radiation from these loops,
though at this stage we can only compare to results for
gravitational radiation\refto{VV}.  Note, however, that the divergences at
small
and large $\phi$ become weakened relative to the previous analysis; this may
reflect a shortcoming of our approximation or the genuine influence of
backreaction. However, a considerably more detailed
quantitative analysis is necessary to test the accuracy of this approach,
especially if we are to normalize it properly for string network
simulations.

A study of the overall Kibble--Turok loop parameter space (which
previous analyses have regarded as fairly typical) yielded an approximate value
$\kappa \approx 0.1$, given the assumed damping coefficient (\dampcoeff) which
was set by normalizing $\Delta$ with the long string results.  However, $\kappa
\approx 0.1$ is the typical backreaction scale expected for GUT-scale global
string loops with $\mu(\Delta) \approx 100f_a^2$ (refer to ref.\refto{VV,BSb}),
thus independently validating our previous normalization \eqnam{\loopnorm} $$
\Delta \approx 0.1L\,.\eqno(\new) $$ Given that the case for the local
backreaction approximation is not as clear-cut for closed loop solutions, these
results must be regarded as encouraging. At the very least,
this approach can be used to phenomenologically incorporate expected loop decay
rates, but results to date suggest it will do substantially better.

\figure{kappavpsi2.eps}{3.5in}{0in}{12}{Parameter dependence of the
backreaction scale $\kappa$ for $\alpha=0$ Kibble--Turok loops using the local
backreaction approximation.}

\sectbegin{4}{Radiative backreaction in an expanding universe}

\noindent Our current understanding of the evolution of a cosmic string network
is based  on a marriage between analytic models and sophisticated network
simulations\refto{BB,ASa}. However, the network simulations only  evolve the
free equations of motion for a string in an expanding universe. In order to
incorporate the radiative effects discussed in the  preceding section one
must modify the equations of motion to include a radiation damping term,
\eqnam{\expandeom}
$$\eqalign{&\mu_0\bigg{[}\ddot{\bf X}+{2\dot a\over a}(1-\dot{\bf
X}^2)\dot{\bf X}-{1\over\epsilon} \bigg{(}{{\bf X}^{\prime}\over
\epsilon}\bigg{)}^{\prime}\bigg{]}={\bf f}\,,\cr
&\mu_0\bigg{[}\dot\epsilon+{2\dot a\over a}
\epsilon\dot{\bf X}^2 \bigg{]}=f^0\,,}\eqno(\new)$$
where $a$ is the scale factor.

However, to calculate this radiation damping term one must use Green
functions in an expanding  background. In the radiation-dominated era,
the retarded Green function is
\eqnam{\greenfn}
$$D_{\rm ret}(x,x^{\prime})={a(\eta)\over 2\pi
a(\eta^{\prime})}\delta\big{(}(x-x^{\prime})^2\big{)}
\theta(\eta-\eta^{\prime})\,.\eqno(\new)$$
where $x^{\mu}=(\eta,{\bf x})$ and $x^{\prime\mu}
=(\eta^{\prime},{\bf x}^{\prime})$. In the matter-dominated era the retarded
Green function also includes the effects of back-scatter off the background
spacetime curvature,
$$D_{\rm ret}(x,x^{\prime})={a(\eta)\over 2\pi
a(\eta^{\prime})}\bigg{[}\delta\big{(}(x-x^{\prime})^2\big{)}
\theta(\eta-\eta^{\prime})+{1\over
2\eta\eta^{\prime}}\theta(\eta-\eta^{\prime}-|{\bf
x}-{\bf x}^{\prime}|)\bigg{]}\,.\eqno(\new)$$
Applying, the `local backreaction approximation' in either of these scenarios,
one finds that the forcing terms are given by,  \eqnam{\newforce}
$$\eqalign{{\bf f}^{\rm rad}&={\bf f}^{\rm rad}_{\rm flat}+{\dot a
\over a}{\bf g}+{\cal O}(1/t^3)\,,\cr
f^{0,{\rm rad}}&=f^{0,{\rm rad}}_{\rm flat}+{\dot a\over a}
g^0+{\cal O}(1/t^3)\,,}$$
where the flat suffix denotes the flat space backreaction force given by
(\approxforce)  and $(g^0,{\bf g})$ is a correction to the force due to
the expanding background.  Notice that to eliminate `runaway' solutions due to
$ {\bf X}^{\tdot}$, we must now resubstitute the damped expanding universe
equations of motion (\expandeom).  Recall that perturbations with lengthscales
$r\hbox{$>\!>$} H^{-1}$ are essentially `frozen' by Hubble damping. It is for
the same reason that large-scale perturbations will not
radiate, despite the high degree of string curvature on these lengthscales.

The forced equations of motion (\expandeom) can be averaged to
derive equations for the evolution of the density of long strings
$(\rho_{\infty})$ and loops $(\rho_{\rm L})$, under the influence of the
expansion, Hubble damping and the radiation backreaction force. If one now
inserts a term to take into account of loop production, the equations become
\eqnam{\scale}
$$\eqalign{&\dot\rho_{\infty}=-{2\dot a\over a}(1+\ll
v^2\rr)\rho_{\infty} - {c\rho_{\infty}\over L} -{d\rho_{\infty}\over L}\,,
\cr &\dot\rho_{\rm L}=-{3\dot a\over
a}\rho_{\rm L}+{c\rho_{\infty}\over L}\,,}\eqno(\new)$$
where
$$d=d_0+{\dot a\over a}d_1+{\cal O}(1/t^2)\,, $$
$\ll v^2\rr$ is the average string velocity, $d_0,d_1,..$ are constants and
$c$ is a measure of the efficiency of loop production.
Substituting $\rho_{\infty}=\mu\zeta/t^2$ and $L=\zeta^{-1/2}t$
into (\scale), one obtains
\eqnam{\eqzeta}
$${\dot\zeta\over\zeta}={1\over t}\bigg{[}2-2\beta(1+\ll
v^2\rr)-(c+d)\zeta^{1/2}\bigg{]}\,,\eqno(\new)$$
where $\beta$ is determined by the scalefactor, $a \propto t^{\beta}$.
Eqn (\eqzeta) has an attractive
fixed point, which corresponds to the scaling regime. If $d_{i}=0$ for $i>0$,
then we have \eqnam{\loopprod}
$$c=\zeta^{-1/2}(1-\ll v^2\rr)-d_0\,.$$
In the case where $d_1$ is non-zero, one should observe transient effects
in the scaling. However, for large times these effects will be negligible and
the attractive fixed point is exactly that for $d_1=0$.

\sectbegin{5}{Conclusions}

\noindent We have introduced, and we have endeavoured to justify, a new
approach to the study of radiation backreaction on strings (and other extended
objects and membranes).  If our analysis and the supporting evidence is valid,
then the `local backreaction approximation' offers  the hope of quantitative
insight into the essentially intractable problem of radiative
damping effects during string network evolution. It is appropriate, therefore,
to
summarize the main points in our discussion.

By exploiting the analogy with classical electrodynamics we have used Green
function methods to separate the self- and radiation fields of a global string,
using the former to renormalize the string tension.  We then approximated
the radiation force at a point on the string by an
expansion in powers of a cut-off parameter $\Delta$.  The `local backreaction
approximation' to the radiation force, then, is the local force at the point in
question multiplied by an effective width $\Delta$ beyond which the
neighbouring string segment contributions become negligible.  We normalize
$\Delta$ in order to reproduce the actual radiation force in known
situations and we have confirmed the self-consistency of the approximation by
demonstrating that $\Delta$ is less than the string curvature radius. We then
generalized this approach to the temporal transverse gauge, a convenient gauge
for studying dissipative string processes. The final step was to remove
unphysical `runaway' solutions, which plague the analogous point-particle
analysis, by resubstituting the string equations of motion.

We then tested a numerical implementation of the `local backreaction
approximation' by investigating a variety of long string and closed loop
trajectories.  We have directly compared the results with analytic radiation
calculations and numerical field theory simulations, demonstrating a consistent
normalization for $\Delta$ using these independent methods.  This approach
reproduces the correct scale-dependence of radiative effects and demonstrates
satisfactory quantitative agreement for a wide variety of different solutions.
There is clearly scope for a more detailed analysis of the accuracy of this
approach and for addressing a number of outstanding issues.  However, we have
presented sufficient grounds for believing that the `local backreaction
approximation' is a significant step forward in the study of string radiative
backreaction.

\nosectbegin{Acknowledgments}

\noindent We are grateful for helpful discussions with Robert Caldwell, Brandon
Carter, Atish Dabholkar and Alex Vilenkin.
The algorithm used for flat space Nambu string evolution was developed by EPS
in
a collaborative project with Bruce Allen (see ref.\refto{ASa}).
We both acknowledge the support of PPARC and EPSRC, in particular the Cambridge
Relativity rolling grant (GR/H71550) and Computational Science Initiative
grants
(GR/H67652 \& GR/H57585).



\def\hang{}


\def\jnl#1#2#3#4#5#6{\hang{#1 [#2], {\it #4\/} {\bf #5}, #6.}
									}


\def\jnlerr#1#2#3#4#5#6#7#8{\hang{#1 [#2], {\it #4\/} {\bf #5}, #6.
{Erratum:} {\it #4\/} {\bf #7}, #8.}
									}


\def\prep#1#2#3#4{\hang{#1 [#2], `#3', #4.}
									}

\def\proc#1#2#3#4#5#6{\hang{#1 [#2], in {\it #4\/}, #5, eds.\ (#6).}
}
\def\procu#1#2#3#4#5#6{\hang{#1 [#2], in {\it #4\/}, #5, ed.\ (#6).}
}

\def\book#1#2#3#4{\hang{#1 [#2], {\it #3\/} (#4).}
									}



\def\prl{Phys.\ Rev.\ Lett.}
\def\pr{Phys.\ Rev.}
\def\pl{Phys.\ Lett.}
\def\np{Nucl.\ Phys.}
\def\prp{Phys.\ Rep.}

\def\cqg{Class.\ Quant.\ Grav.}

\def\mn{M.$\,$N.$\,$R.$\,$A.$\,$S.}

\def\cup{Cambridge University Press}

\def\skip{\vskip -4pt}

\nosectbegin{References}

\references

\baselineskip 16pt
\let\it=\nineit
\let\rm=\ninerm
\let\bf=\ninebf
\rm

\refis{BSa}
\jnl{Battye, R.A., \& Shellard, E.P.S.}{1994}{Global string
radiation}{\np}{B423}{260}

\refis{BSb}
\jnl{Battye, R.A., \& Shellard, E.P.S.}{1994}{Axion string
constraints}{\prl}{73}{2954}

\refis{BSc}
\prep{Battye, R.A., \& Shellard, E.P.S.}{1994}{String radiative backreaction
}{DAMTP Preprint R94/31, to appear {\it\prl}}

\refis{BSd}
\prep{Battye, R.A., \& Shellard, E.P.S.}{1995}{Non-linear interactions on
global topological defects}{DAMTP Preprint}

\refis{Bata}
\prep{Battye, R.A. \& Carter, B.}{1995}{Gravitational perturbations of
relativistic membranes and strings}{DAMTP
Preprint R95/17, to appear {\it\pl} {\bf B}}
\prep{Battye, R.A.}{1995}{Gravitational radiation and backreaction for
cosmic strings}{DAMTP Preprint R95/18, submitted to {\it\cqg}}

\refis{ASa}
\jnl{Allen, B., \& Shellard, E.P.S.}{1990}{Cosmic string evolution---a
numerical simulation}{\prl}{64}{119} \proc{Shellard, E.P.S., \& Allen,
B.}{1990}{On the evolution of
cosmic strings}{Formation and Evolution of Cosmic Strings}{Gibbons, G.W.,
Hawking, S.W., \& Vachaspati, T.}{\cup} \skip

\refis{AC}
\jnl{Bennett, D.P., \& Bouchet, F.R.}{1991}
{Constraints on the gravity wave background generated by cosmic
strings}{\pr}{D43}{2733}
\jnl{Caldwell, R.R., \& Allen, B.}{1992}{Cosmological constraints on cosmic
string gravitational radiation}{\pr}{D45}{3447}

\refis{KR}
\jnl{Kalb, M., \& Ramond, P.}{1974}{Classical direct interstring
action}{\pr}{D9}{2273}\jnl{Witten E.}{1985}{Cosmic
Superstrings}{\pl}{153B}{243}

\refis{VV}
\jnl{Vilenkin, A., \& Vachaspati, T.}{1987}{Radiation of Goldstone bosons from
cosmic strings}{\pr}{D35}{1138}
{\it See also} \jnl{Vachaspati, T., \& Vilenkin, A.}{1985}{Gravitational
radiation from cosmic
strings}{\pr}{D31}{3052}

\refis{DSb}
\jnl{Davis, R.L., \& Shellard, E.P.S.}{1988}{Antisymmetric tensors and
spontaneous symmetry breaking}{\pl}{214B}{219}

\refis{DQ}
\jnl{Dabholkar, A., \& Quashnock, J.M.}{1990}{Pinning down the
axion}{\np}{B333}{815}

\refis{HCH}
\jnl{Copeland, E., Haws, D., \& Hindmarsh, M.B.}{1990}{Classical theory of
radiating
strings}{\pr}{D42}{726}

\refis{dirac}
\jnl{Dirac, P.A.M}{1938}{}{Proc. Roy. Soc. London}{A167}{148}
\book{Barut, A.O.}{1964}{Electrodynamics and classical theory of fields
and particles}{Macmillan, New York}
\book{Jackson, J.D.}{1975}{Classical Electrodynamics}{Wiley, New York}

\refis{GV}
\jnl{Garfinkle, D., \& Vachaspati, T.}{1987}{Radiation from kinky, cuspless,
cosmic
loops}{\pr}{D36}{2229}

\refis{kibble}
\jnl{Kibble, T.W.B.}{1980}{Some implications of a cosmological phase
transition}{\prp}{67}{183}

\refis{kibbturok}
\jnl{Kibble, T.W.B., \& Turok, N.G.}{1982}{Selfintersection of cosmic
strings}{\pl}{116B}{141}

\refis{QS}
\jnl{Quashnock, J.M. \& Spergel, D.N.}{1990}{Gravitational self-interactions of
cosmic
strings}{\pr}{D42}{2505}

\refis{sod}
\book{Sod, G.A.}{1985}{Numerical methods in fluid dynamics}{\cup}

\refis{Sak}
\jnl{Sakellariadou, M.}{1990}{Gravitational waves emitted from infinite
strings}{\pr}{D42}{354}
\jnl{Sakellariadou, M.}{1991}{Radiation of Nambu--Goldstone bosons from
infinitely long cosmic strings}{\pr}{D44}{3767}

\refis{Vac}
\jnl{Vachaspati, T.}{1986}{Gravitational effects of
cosmic strings}{\np}{B277}{593}  \skip

\refis{VacFort}
\jnl{Fort, J. \& Vachaspati, T.}{1993}{Do global strings form black
holes}{\pl}{311B}{41}

\refis{Hina}
\jnl{Hindmarsh, M.B.}{1990}{Gravitational radiation from kinky
infinite strings}{\pl}{251B}{28}

\refis{DavVV}
\jnl{Vilenkin, A., \& Everett, A.E.}{1982}{Cosmic strings and domain walls in
models with Goldstone and pseudo-Goldstone bosons}{\prl}{48}{1867} \jnl{Davis,
R.L.}{1985}{Goldstone bosons in string models of galaxy
formation}{\pr}{D32}{3172}

\refis{Sd}
\procu{Shellard, E.P.S.}{1992}{The numerical study of topological
defects}{Approaches
to Numerical Relativity}{d'Inverno, R.}{\cup}

\refis{structure}
\jnl{Zel'dovich, Ya.B.}{1980}{Cosmological fluctuations produced
near a singularity}{\mn}{192}{663}
\jnlerr{Vilenkin, A.}{1981}{Cosmological density fluctuations produced by
vacuum
strings}{\prl}{46}{1169}{46}{1496}

\refis{BB}
\jnl{Bennett, D.P., \& Bouchet, F.R.}{1990}{High resolution simulations of
cosmic
string evolution: network evolution}{\pr}{D41}{2408} {\it See also}
\jnl{Albrecht, A., \& Turok, N.}{1989}{Evolution of cosmic string
networks}{\pr}{D40}{973}

\endreferences

\vfill
\end


In this
approximation the string appears to be straight at long distances from the
core,
corresponding to the physical long range fields being uncorrelated beyond some
distance $\Delta$. If the radiation backreaction force calculated in this
procedure is incapable of explaining the actual strength of radiation, then the
assumptions on which it is based must be incorrect. Conversely, if a particular
value of $\Delta$, less than the curvature radius, is sufficient to explain the
strength of Goldstone radiation in a realistic context, then the local
backreaction approximation  will be valid for sensible string
trajectories.
Possibly the easiest way to understand the construction of the radiation
backreaction force, in the local backreaction approximation, is by considering
fig.~3. Fig.~3(a) shows a schematic of how the actual radiation
backreaction force depends on contributions from the points on the string a
distance $x$ from the point in question. The precise rate of fall off of the
curve is unknown, however it must be faster than $1/R$ for the sum of the
contributions to be finite. The total area under the curve represents the exact
magnitude of the radiation backreaction force.  Fig.~3(b) represents the
equivalent to fig.~3(a) for the local backreaction approximation. Effectively
we calculate the backreaction force at the point in question and assume that
the contribution from surrounding points is exactly the same and impose a
cut off at $\Delta$. If one can find a value of
$\Delta$, less than the radius of curvature, for which the area under the two
curves is equal, then the approximation is valid.

\figure{small.eps}{2.6in}{-0.35in}{4}{A schematic showing how the fixed
integration region used in the 'local backreaction approximation' breaks
relativistic covariance on the worldsheet. In (a) we show the domain of
dependence in the 'local backreaction approximation' and in (b) we show an
equivalant region of the worldsheet under a lorentz boost. We argue that this
will only have an effect on the backreaction force for perverse string
trajectories.}